\documentclass[12pt,reprint, aps, prd, showpacs,nofootinbib]{revtex4}

\usepackage[utf8]{inputenc}
\usepackage[T1]{fontenc}
\usepackage{amsmath,amssymb}
\usepackage{hyperref}
\usepackage{graphicx}
\usepackage{subfigure}

\usepackage[utf8]{inputenc}
\usepackage[T1]{fontenc}
\usepackage{amsmath,amssymb}

\DeclareMathOperator{\sgn}{sgn}

\usepackage{xcolor}

\begin{document}

\title{Cosmological perturbations in Hybrid Loop Quantum Cosmology: Mukhanov-Sasaki variables}

\pacs{04.60.Pp, 04.60.Kz, 98.80.Qc }

\author{Laura Castell\'o Gomar}
\email{laura.castello@iem.cfmac.csic.es}
\affiliation{Instituto de Estructura de la Materia, IEM-CSIC, Serrano 121, 28006 Madrid, Spain}
\author{Mikel Fern\'andez-M\'endez}
\email{m.fernandez.m@csic.es}
\affiliation{Instituto de Estructura de la Materia, IEM-CSIC, Serrano 121, 28006 Madrid, Spain}
\author{Guillermo A. Mena Marug\'an}
\email{mena@iem.cfmac.csic.es}
\affiliation{Instituto de Estructura de la Materia, IEM-CSIC, Serrano 121, 28006 Madrid, Spain}
\author{Javier Olmedo}
\email{jolmedo@fisica.edu.uy}
\affiliation{Instituto de F\'{i}sica, Facultad de Ciencias, Igu\'a 4225, esq.\ Mataojo, Montevideo, Uruguay}

\begin{abstract}
We study cosmological perturbations in the framework of Loop Quantum Cosmology, using a hybrid quantization approach and Mukhanov-Sasaki variables. The formulation in terms of these gauge invariants allows one to clarify the independence of the results on choices of gauge and facilitates the comparison with other approaches proposed to deal with cosmological perturbations in the context of Loop Quantum Theory. A kind of Born-Oppenheimer ansatz is employed to extract the dynamics of the inhomogeneous perturbations, separating them from the degrees of freedom of the Friedmann-Robertson-Walker geometry. With this ansatz, we derive an approximate Schr\"odinger equation for the cosmological perturbations and study its range of validity. We also prove that, with an alternate factor ordering, the dynamics deduced for the perturbations is similar to the one found in the so-called {\it dressed metric approach}, apart from a possible scaling of the matter field in order to preserve its unitary evolution in the regime of Quantum Field Theory in a curved background and some quantization prescription issues. Finally, we obtain the effective equations that are naturally associated with the Mukhanov-Sasaki variables, both with and without introducing the Born-Oppenheimer ansatz, and with the different factor orderings that we have studied.
\end{abstract}

\keywords{Quantum cosmology, Loop Quantum Gravity, cosmological perturbation theory}

\maketitle

\section{Introduction}

Since the pioneer work by Lifshitz \cite{Lpert}, the study of perturbations has played a prominent role in cosmology \cite{perturbations,bardeen,theorypertu,mukhanov}. In a rough approximation, our Universe seems to be homogeneous and isotropic at sufficiently large scales, described by what is usually called a Friedmann-Robertson-Walker (FRW) spacetime. This approximation is supported not only by a combination of observations and basic assumptions, but also by some theoretical results \cite{egs}, at least for certain matter contents. This  homogeneity and isotropy (in a suitable average) leads to the question of how the structures superposed to it formed and developed. The theory of cosmological perturbations \cite{theorypertu} together with the paradigm of inflation \cite{inflation} provide a remarkably successful explanation. This explanation is valid both for the formation of large scale structures and for the fine details of the cosmic background radiation. The measurement of the fluctuations of this primordial radiation, which originated in the small perturbations that were present in the Early Universe, is a central core of what is nowadays called precision cosmology, an era in which technology has allowed such a good observation of cosmological phenomena in astronomy and astrophysics as to make possible for the first time the determination of a number of the most important cosmological parameters with several digits of significance \cite{precision}. The last episode has been the observation of the BB-spectrum of the cosmic radiation by BICEPS2 \cite{biceps}, which seems to confirm the predictions based on tensor perturbations in inflationary cosmology.

Although perturbations in cosmology admit a classical formulation, and in fact it is remarkable how well this classical treatment is capable of predicting the present observations, the very nature of the perturbations is rather quantum mechanical. In the predictions of the primordial power spectrum, Quantum Field Theory (QFT) in a curved background already enters at a certain level in order to explain in a natural way the (at least almost) Gaussian distribution of the primordial fluctuations in the Early Universe \cite{mukhanov}. For this, essentially, one describes the perturbations by quantum fields and assumes that they are initially in a vacuum state with the maximal symmetry of a de Sitter spacetime (a Bunch-Davis state \cite{bunchdavies}), which describes rather well the inflationary stage of the Universe. Techniques of QFT in curved spacetimes can then be employed to analyze and regularize the contributions of these quantum fields on the fixed cosmological background. The ultimate hope of the community of physicists working in the quantization of gravity, nonetheless, is that the relics of the quantum fluctuations of the Early Universe may encode information about the quantum nature of the spacetime geometry itself. In this way, rather than considering QFT in fixed cosmological backgrounds as the last step in the progress of understanding the primordial fluctuations of our Universe, and viewing the quantum fields of the perturbations exclusively as test fields that propagate in a given geometry (which can be purely classical, but may also be quantum mechanically corrected), one would hope for a quantum theory which incorporates both the geometry and the perturbations, with interplay between them, and which is potentially predictive. At the end of the day, the goal would be identifying windows for the observation of traces of the Early Universe phenomena, in order to detect any of those predictions and falsify the model, or even the theory of quantum gravity from which it has been derived (provided that this derivation is not based in other extra assumptions and is therefore essentially unique). In particular, of course, only when the homogeneous background and the inhomogeneous perturbations are treated quantum mechanically on a similar footing, it is possible to speak about a quantum structure for a geometry that includes those background and perturbations.

In this context, a lot of attention has been devoted lately to develop a formalism for cosmological perturbations in the framework of Loop Quantum Cosmology (LQC). LQC \cite{lqc,lqc2} is the study of cosmological systems with the methods of Loop Quantum Gravity (LQG) \cite{lqg}, a nonperturbative and background independent program for the quantization of general relativity that provides nowadays one of the most appealing candidates for a quantum theory of the gravitational interaction. LQC has been applied successfully to homogeneous scenarios in cosmology, not only isotropic FRW ones with various kinds of matter content \cite{APS1,APS2,MMO,MOP,FRWLQC}, but also anisotropic models of different Bianchi types \cite{BianchiLQC}. One of the most remarkable predictions is the resolution of the Big Bang singularity, which is unavoidable in the classical Einstein theory (see, e.g., \cite{HEllis}), and which is replaced by a turnover called Big Bounce at least in some specific families of states with a marked classical behavior \cite{APS1,APS2,taveras}. The limitation of homogeneity is a clear restriction in this quantum treatment of the geometry and of the spacetime structure in cosmology; therefore, it is natural to try and go beyond the assumption of homogeneity in the analysis of cosmological universes. Cosmological perturbations are an optimal arena for that, both because of the level of understanding and development of their classical treatment and because of their physical relevance.

Two main lines of attack have been followed in this analysis within LQC. One of the approaches provides a scheme to derive effective equations for the perturbations which capture the effects of the quantum nature of the spacetime geometry \cite{effective,cai,bojogianlufren}. The approach is based on the need that the algebra of constraints closes in the quantum theory. This restricts the possible quantum corrections to the constraints of general relativity. Together with assumptions about the corrections expected in LQG (coming from the use of holonomies and the regularization of the inverse of the volume operator), a series of technical (and less obvious) hypotheses (about validity of expansions, choice and range of canonical variables, locality, etc.), and the introduction of a structure of Poisson brackets for the expectation values and moments of the basic variables, this scheme allows one to study the modified field equations for the perturbations. The other line of attack deals
with the direct quantization of the FRW geometry and the perturbations \cite{hybridsphere,hybridflat,ampli,AAN,AAN1,AAN2}. In principle, both types of approaches are complementary, since some of the assumptions used in the derivation of effective equations from the closure of the algebra would ultimately be possible to check only when one has at his disposal a genuine quantum treatment. On the other hand, to extract physical predictions from the genuine quantum description, one needs to understand the effective regimes that are consistent with the fundamental symmetries and properties of the system.

The works confronting the quantum description of FRW universes with perturbations try and combine a genuine loop quantization of the FRW geometry with a homogeneous matter content together with a more conventional Fock quantization of the perturbations of the geometry and matter fields \cite{hybridsphere,hybridflat,AAN1,AAN2}. The idea is inspired in the hybrid approach to LQC that was originally developed in the first inhomogeneous cosmologies quantized to completion in the framework of the loop formulation, namely, the Gowdy models with linear polarization of the gravitational waves \cite{hybridgowdy}. Gowdy cosmologies are spacetimes with two spatial Killing vectors and compact spatial sections, which can only be homeomorphic to a three-torus, a three-sphere, or a three-handle \cite{gowdy}. In the case in which the inhomogeneous degrees of freedom of the metric describe only one of the two possible polarizations of the gravitational waves (more explicitly, waves with linear polarization), these cosmologies have been quantized within an exact treatment of the geometry without the need of a perturbative truncation, even in the presence of matter scalar fields \cite{gowdymatt}. The Fock quantization of the inhomogeneous modes of the metric and matter fields was picked out in Refs. \cite{gowdytime,gowdyunique} by demanding criteria of invariance under the spatial isometries of the model and the unitary implementability of the dynamics. Actually, these criteria proved to select a unique canonical pair to describe the inhomogeneous fields among all the pairs that are related by a scaling of the field configuration by a function of the homogeneous (background) geometry \cite{gowdytime}. Besides, the same criteria select a unique class of unitarily equivalent Fock representations for the commutation relations of the privileged canonical pair \cite{gowdyunique}.

In a similar manner, these criteria can be applied as well to choose a unique Fock quantization of the inhomogeneities in more general scenarios than the Gowdy cosmologies \cite{uniqueness1,uniqueness2,uniquenessother,uniqueperturb}. For instance, following the hybrid approach, these uniqueness criteria guided the quantization of perturbations around FRW spacetimes in Refs. \cite{hybridsphere,hybridflat} (specialized to the case of spherical and of compact flat spatial topologies). That hybrid quantization rested, essentially, on two assumptions. First, as we have mentioned, it rested on the hypothesis that the most relevant effects of the loop quantum geometry are those that affect the zero modes which describe the degrees of freedom of the FRW geometry, so that one can adopt a hierarchy in the quantization where the other geometry degrees of freedom admit a more conventional, quantum Fock formulation. Second, it rested on the truncation of the system at quadratic perturbative order in the action, considering the inhomogeneities in the matter field and the metric as linear perturbations, and splitting them from the homogeneous, zero modes of the system. A recent discussion about how this truncation allows for a consistent symplectic description can be found in Ref. \cite{bojogianlufren}. This is rather straightforward if one starts with the gravitational action written in Hamiltonian form. It suffices to substitute in that action the expressions of the gravitational and matter variables in terms of zero modes and inhomogeneous perturbations, and truncate the result at quadratic order. By construction, one obtains a symplectic structure for the system containing perturbations, as well as the constraints to which this system is subject, arising from those of the gravitational theory at the order of truncation adopted in the action.

Though mathematically this truncation is clearly consistent, there has been some confusion about it and its physical interpretation. For instance, it has been claimed \cite{AAN1,AAN2} that one has to renounce to a symplectic description of the perturbed FRW universes. The price to be paid then is that the perturbations must be viewed {\emph{just}} as test fields of a {\it{dressed}} FRW geometry (which incorporates LQC effects), and hence one has to abandon a genuine quantum description of the geometry including perturbations, developing instead an extension of QFT in curved backgrounds to the dressed metric scenario. In doing so, one also ought to renounce to the possibility of defining quantum metric operators beyond the homogeneous and isotropic truncation of the studied cosmologies. The confusion seems to originate from the fact that a perturbative truncation of a given order in the action \cite{hybridsphere,hybridflat,HH,kiefer} does not correspond to the same order of perturbative truncation in all the metric (and matter) degrees of freedom of the system, owing to the nonlinearity of the equations of general relativity (a recent discussion about this fact is addressed also in Ref. \cite{bojogianlufren}). Again, the experience gained with the analysis of the Gowdy cosmologies is extremely valuable to clarify the situation. In the (almost) gauge fixed model for the case of three-torus spatial topology, the inhomogeneous degrees of freedom can be described by a metric field with no zero mode that satisfies a linear second-order equation of Klein-Gordon type (on an auxiliary space identifiable as the circle \cite{gowdylargo}) with no sources. We can expand this field in a perturbative series. The linearity of the field equation implies that the solution for the $n$-th power contribution to the field in this perturbative expansion is itself, by its own, an exact solution. In other words, different perturbative orders decouple in the field equation. With any of these solutions (modulo a momentum and a Hamiltonian global constraints and together with a solution for the zero modes of the model) one can construct an exact solution for the spacetime metric. The formulas can be found in the Appendix of Ref. \cite{gowdylargo}. It is straightforward to check that the metric gets contributions of perturbative orders different from those of the considered field solution. For instance, if one considers a solution of linear perturbative order in the expansion of the inhomogeneous field, the metric gets perturbative corrections of all orders. Even if one focuses the attention on metric components (something that is meaningful in the gauge fixed system) and considers logarithms of the diagonal ones, it is easy to see that these metric quantities get contributions beyond the linear perturbative approximation. Obviously, nevertheless, nothing is inconsistent in the description and treatment of the system, and in particular in its consideration as a constrained symplectic one. One of the goals of the present work is to show how one can construct a formalism for cosmological perturbations around FRW that can be considered similar to that proposed in Refs. \cite{AAN1,AAN2} but without abandoning the view that the quantum theory describes a constrained manifold supplied with a symplectic structure, as it is the standard case in gravity. In such a formalism, hence, one can face questions about the genuine quantum nature of the perturbed geometry and the associated spacetime structure.

In the previous analysis of cosmological perturbations using the hybrid approach to LQC, variables adapted to gauge fixed reductions of the system were employed \cite{hybridsphere,hybridflat}. This has several drawbacks. First, it leads to the wrong impression that the results are intrinsically gauge dependent. Although it was proven in Refs. \cite{hybridsphere,hybridflat,uniqueperturb} that, in the regime in which the inhomogeneities admit a description by means of a QFT in a curved background (which includes LQC modifications with respect to general relativity), this QFT is unitarily equivalent to one based on annihilation and creation-like variables constructed from gauge invariants, the discussion of the formalism is obscured by the use of variables which are not invariant in fact. The introduction of gauge invariants makes easier to discern the extent to which the approach restricts the classical and quantum freedom in the gauge transformations of the perturbed system. In particular, in the flat case, one would like to describe the perturbations in terms of Mukhanov-Sasaki (MS) variables \cite{MSvariable}. On the one hand, these variables are perturbative gauge invariants and allow an almost straightforward discussion of the primordial power spectrum, because their spectrum is related in a simple way to that of the co-moving curvature perturbations. Besides, they satisfy a Klein-Gordon equation in an auxiliary static spacetime with a time dependent quadratic potential. Remarkably, it is precisely for this kind of equations that our criteria of spatial symmetry invariance and unitary dynamics can be directly applied to pick out a unique Fock quantization. On the other hand, the use of MS variables permits the comparison of the hybrid approach with other proposals for the treatment of cosmological perturbations in LQC, and specifically with the dressed metric proposal, since the latter has been expressed in terms of these gauge invariants \cite{AAN1,AAN2}. Finally, the formulation in terms of MS variables can be regarded as a previous step towards the introduction of a canonical transformation in the system aimed at describing the inhomogeneous perturbations by these variables, the linear perturbative (gauge) constraints, and their corresponding momenta. Completing this transformation into a canonical one in the entire system, including zero modes, one can obtain a quantum theory in the hybrid approach where the gauge dependence is fully understood \cite{inpreparation}. In this manner, the formulation in terms of MS variables sheds light on some recent discussion about the role of gauge fixation in the separation of zero modes from inhomogeneous perturbations in the hybrid approach. Actually, this separation makes use of the mode expansion associated with the Laplace-Beltrami operator of the spatial sections, for whose construction one needs just an auxiliary spatial metric already available in the FRW system.

In the rest of this work, therefore, we will present the hybrid quantization in terms of MS variables of perturbed flat FRW universes with compact spatial sections in the presence of a matter scalar field. The basic results and formulas of previous studies of this system in the literature will be summarized in Sec. 2, where we will also introduce the change of variables for the inhomogeneous modes that leads to the MS invariants. This change will be completed into a canonical transformation for the perturbed FRW model in Sec. 3. In that section, we will also derive the expression of the quadratic contribution of the inhomogeneous perturbations to (the zero mode of) the Hamiltonian constraint in terms of the introduced MS variables, showing that it reproduces the so-called MS Hamiltonian for a proper scaling of the inhomogeneities. We will quantize this constrained system in Sec. 4, following the hybrid approach. In Sec. 5, we will adopt a kind of Born-Oppenheimer (BO) ansatz for the quantum states. With that ansatz, and neglecting nondiagonal terms in the homogeneous (FRW) quantum geometry, we will be capable to pass from (the zero mode of) the Hamiltonian constraint to a Schr\"odinger equation in the internal time provided by the homogeneous part of the matter field. We will also compare this Schr\"odinger equation with that put forward in Refs. \cite{AAN1,AAN2} by ``deparametrizing'' the system and employing the dressed metric QFT approach. Next, in Sec. 6, we will introduce a different factor ordering for the quantization of our constrained and symplectic system. We will show that this factor ordering, again after using  a BO ansatz and ignoring nondiagonal elements in the homogeneous geometry, leads to a quantum equation for the propagation of the inhomogeneous perturbations which is similar to that of Agull\'o, Ashtekar, and Nelson. Essentially, the differences refer to the choice of scaling for the inhomogeneous field that is quantized \`a la Fock, and to possible ambiguities in the operator representations selected in the quantization. Since, in the light of this result (and leaving aside the scaling of the inhomogeneous modes), the main discrepancy between our hybrid construction and the construction of Refs. \cite{AAN1,AAN2} may be interpreted as an alternate choice in factor ordering, it will be then easy to identify the difference between the  corresponding quantum propagation equations for the inhomogeneous perturbations. In Sec. 7, we will compute and compare the effective equations for the MS invariants that follow from our hybrid approach using the quantum prescriptions of Ref. \cite{hybridflat}, on the one hand, and with the alternate factor ordering that can be related to the dressed metric approach, on the other hand. Finally, we will conclude in Sec. 8.

\section{Perturbed FRW universes: The system}\label{sec:system}

In this section, we will provide a summary of the classical description of our cosmological system. This classical model will be the starting point for our quantum analysis, in which we will combine mathematical tools of LQC and Fock quantization techniques. Most of the details and formulas can be found in Ref. \cite{hybridflat}. Thus, we are interested in studying inhomogeneous perturbations of FRW spacetimes with compact flat spatial sections and a matter content given by a minimally coupled scalar field. We will focus our attention on the case in which this field $\Phi$ is subject to a potential that consists only of a mass term. The extension of our analysis to other potentials is almost straightforward. On the other hand, we will consider exclusively scalar perturbations of the geometry. This is fully consistent, since these perturbations decouple (at our truncation perturbative order) from other kinds of perturbations (namely, vectors and tensor perturbations \cite{bardeen}). In fact, the study of the physical degrees of freedom included in the tensor perturbations can be carried out in a completely similar way, and is actually simpler from a technical point of view.

We adopt a 3+1 decomposition of the metric in Arnowitt-Deser-Misner (ADM) form (see, e.g., \cite{mtw}), expressing it in terms of the three-metric $h_{ij}$ induced on the sections of constant time $t$, a lapse function $N$, and a shift vector $N^i$ (or covector $N_i$). Spatial indices $i,j$ run from 1 to 3. In an FRW spacetime, these metric functions are completely characterized by a homogeneous lapse $N_0(t)$, the logarithm of the scale factor of the spatial metric $\alpha(t)$, and a static auxiliary three-metric $^0h_{ij}$. In the considered case of compact flat universes, we can take $^0h_{ij}$ as the standard flat metric on the three-torus $T^3$, with period equal to $l_0$ in each of the orthonormal directions, for which we choose angular coordinates $\theta_i$ such that $2\pi \theta_i/ l_0 \in S^1$. Using the auxiliary metric $^0h_{ij}$ (or rather the line element ${}^0h_{ij}d\theta_i d\theta_j$), we can define a volume element on the spatial sections, construct the Hilbert space of functions on those sections that are square integrable with respect to that volume element, and introduce in that space the Laplace-Beltrami operator compatible with the metric  $^0h_{ij}$. The eigenmodes of this operator provide a basis on the considered Hilbert space of functions. Hence, any function in it can be expanded in those modes. In particular, we can expand our inhomogeneous perturbations, transforming the problem of studying the spatial dependence into a spectral analysis in terms of such modes.

In the compact flat case considered here, we can adopt a basis of real Fourier modes, formed by the sine and cosine functions
\begin{equation}
\tilde Q_{\vec n,+} (\vec\theta)= \sqrt 2\cos\left(\frac{2\pi}{l_0}\vec n\cdot\vec\theta\right),\quad  \tilde Q_{\vec n,-} (\vec\theta)= \sqrt 2\sin\left(\frac{2\pi}{l_0}\vec n\cdot\vec\theta\right),
\end{equation}
where $\vec n=(n_1,n_2,n_3)\in\mathbb Z^3$ is any tuple whose first nonvanishing component is a strictly positive integer (in order to avoid repetition of modes). Besides, we have used the notation $\vec n\cdot\vec\theta=\sum_in_i\theta_i$. These modes have a norm equal to the square root of the auxiliary volume $l_0^3$ of the three-torus, and their Laplace-Beltrami eigenvalue is $-\omega_n^2=-4\pi^2\vec n\cdot\vec n/l_0^{2}$. Furthermore, since our inhomogeneous perturbations have {\emph{no zero mode}} contributions, the value $\vec n =  0$ is excluded in the expansion of the inhomogeneities.

Employing this Fourier expansion, the ADM metric can be written as
\begin{eqnarray}\label{eqs:expansions}
h_{ij}(t,\vec\theta) &=& \sigma^2 e^{2\alpha(t)}\;{}^0h_{ij}(\vec\theta)\left[1+2\sum_{\vec n,\epsilon}a_{\vec n,\epsilon} (t)\tilde Q_{\vec n,\epsilon}(\vec\theta)\right] \nonumber\\
&+&6 \sigma^2 e^{2\alpha(t)}\sum_{\vec n,\epsilon}b_{\vec n,\epsilon}(t)\left[\frac1{\omega_n ^2}(\tilde Q_{\vec n,\epsilon})_{|ij}(\vec\theta)+\frac13{}^0h_{ij}(\vec\theta)\tilde Q_{\vec n,\epsilon}(\vec\theta)\right],\\
\label{lapse}
N(t,\vec\theta) &=& \sigma N_0(t)\left[1+\sum_{\vec n,\epsilon}g_{\vec n,\epsilon}(t)\tilde Q_{\vec n,\epsilon}(\vec\theta)\right], \\ \label{eqs:expansions2}
N_i(t,\vec\theta) &=& \sigma^2e^{\alpha(t)}\sum_{\vec n,\epsilon}\frac1{\omega_n^2}k_{\vec n,\epsilon}(t)(\tilde Q_{\vec n,\epsilon})_{|i}(\vec\theta),
\end{eqnarray}
and the scalar field as
\begin{equation}
\Phi(t,\vec\theta) = \frac{1}{\sigma\sqrt{l_0^{3}}}\left[\varphi(t)+\sum_{\vec n,\epsilon}f_{\vec n,\epsilon}(t)\tilde Q_{\vec n,\epsilon} (\vec\theta)\right].
\end{equation}
Here, $\sigma^2=4\pi G/(3l_0^3)$, $G$ is the Newton constant, the vertical bar stands for the covariant derivative with respect to the auxiliary metric ${}^0h_{ij}$, and $\epsilon=+,-$ (for cosine and sine modes, respectively). As we have already commented, in all the sums over the tuples $\vec n$ the zero mode is eliminated. This mode is accounted for by considering the homogeneous metric and field variables, where we include its contribution. The variable $\varphi$ is the homogeneous part of the field. The time dependent Fourier coefficients in these expansions parametrize the inhomogeneities.

Substituting these formulas in the Hamiltonian form of the gravitational action coupled to the scalar field, and truncating the result at quadratic order in the inhomogeneous perturbations, one obtains (in addition to the Legendre term containing the information about the symplectic structure of the system) a total Hamiltonian $H$ which is a linear combination of constraints, with the form \cite{hybridflat,HH}
\begin{equation}\label{eq:Hamiltonian}
H = N_0\Big[H_{|0}+\sum_{\vec n,\epsilon}H^{\vec n,\epsilon}_{|2}\Big]+\sum_{\vec n,\epsilon}N_0g_{\vec n,\epsilon}H^{\vec n,\epsilon}_{|1}+\sum_{\vec n,\epsilon}k_{\vec n,\epsilon}H^{\vec n,\epsilon}_{\_1},
\end{equation}
where $H^{\vec n,\epsilon}_{|1}$ and $H^{\vec n,\epsilon}_{\_1}$ are linear in the inhomogeneous perturbations (we will refer to them as the linear perturbative constraints) and arise from the perturbation, respectively, of the Hamiltonian and the momentum constraints that generate, also respectively, time reparametrizations and spatial diffeomorphisms in general relativity. On the other hand, $H^{\vec n,\epsilon}_{|2}$ is quadratic in the perturbations, and provides the contribution of the inhomogeneities to the zero mode of the Hamiltonian constraint, which in the unperturbed case is just
\begin{equation}\label{eq:H_0}
H_{|0} = \frac{e^{-3\alpha}}{2}\big(-\pi_\alpha^2+\pi_\varphi^2+e^{6\alpha}\bar m^2\varphi^2).
\end{equation}
The constant $\bar m$ is related to the mass $m$ of the scalar field by $\bar m= m \sigma$, and we have called generically $\pi_q$ the momentum conjugate to the variable $q$.

Following the analysis of Ref. \cite{hybridflat}, we introduce now a convenient gauge fixing for the system, though later on we will reformulate our description in terms of gauge invariants. The gauge fixation simplifies the discussion considerably. The adoption of gauge invariants should remove any dependence on the choice of gauge. Actually, gauge invariants are defined as variables which commute with the linear perturbative constraints. One can then search for a set of variables for the inhomogeneous perturbations consisting of the gauge invariants, the mentioned constraints, and suitable momenta for them that might be used as variables whose value can be fixed to remove the gauge freedom. By completing this change of variables for the perturbations into a canonical transformation for the entire system, including zero modes, one would reach a description that is genuinely independent of the (perturbative) gauge, in which the physical degrees of freedom are straightforward to identify. We assume that gauge fixing and the adoption of gauge invariants to describe the perturbations are processes that commute; we will provide a detailed discussion of the system with the outlined strategy without gauge fixing in a future work \cite{inpreparation}. As we will see, the procedure presented here is most convenient to cope with the calculations and compare the hybrid approach with other quantization approaches for cosmological perturbations, like that in LQC of Refs. \cite{AAN1,AAN2}.

We thus adopt a longitudinal gauge, picked out by the pairs of conditions $b_{\vec n,\epsilon} = 0$ and $\pi_{a_{\vec n,\epsilon}}-\pi_\alpha a_{\vec n,\epsilon}-3\pi_\varphi f_{\vec n,\epsilon} = 0$ \cite{hybridflat}, which remove the gauge freedom associated with the linear perturbative constraints. In this gauge, the shift vector vanishes, and the spatial metric is conformal to the flat one. The reduced system obtained with these conditions is subject only to one constraint, namely, the zero mode of the Hamiltonian constraint, and admits a symplectic structure, induced from that of the original system at our order of quadratic truncation in the action, which makes the following a canonical set of phase space variables
\begin{subequations}\label{eqs:newvariables}
\begin{align}
\bar f_{\vec n,\epsilon} &= e^\alpha f_{\vec n,\epsilon}, \\
\pi_{\bar f_{\vec n,\epsilon}} &= e^{-\alpha}(\pi_{f_{\vec n,\epsilon}}-3\pi_\varphi a_{\vec n,\epsilon}-\pi_\alpha f_{\vec n,\epsilon}), \\
\bar\alpha &= \alpha+\frac12\sum_{\vec n,\epsilon}\big(a_{\vec n,\epsilon}^2+f_{\vec n,\epsilon}^2\big), \\
\pi_{\bar\alpha} &= \pi_\alpha +\sum_{\vec n,\epsilon}\big(\pi_\alpha f_{\vec n,\epsilon}^2+3\pi_\varphi a_{\vec n,\epsilon}f_{\vec n,\epsilon}
-f_{\vec n,\epsilon}\pi_{f_{\vec n,\epsilon}}\big), \\
\bar\varphi &= \varphi+3\sum_{\vec n,\epsilon}a_{\vec n,\epsilon}f_{\vec n,\epsilon}, \\
\pi_{\bar\varphi} &= \pi_\varphi,
\end{align}
\end{subequations}
where
\begin{equation}\label{eq:a_n}
a_{\vec n,\epsilon} = 3\frac{\pi_\varphi\pi_{f_{\vec n,\epsilon}}+\big(e^{6\alpha} \bar m^2\varphi-3\pi_\alpha\pi_\varphi\big)f_{\vec n,\epsilon}}{9\pi_\varphi^2+\omega_n^2e^{4\alpha}}.
\end{equation}
Note that the new barred variables for the homogeneous degrees of freedom get quadratic contributions from the inhomogeneities in order to maintain the system symplectic. Thus, if one expresses the metric in terms of these variables, the zero mode part of the metric will get a quadratic perturbative contribution, which nonetheless is not independent of the linear perturbations in the inhomogeneous modes.

The only remaining constraint, as we have said, is $H = N_0[H_{|0}+\sum_{\vec n,\epsilon}H^{\vec n,\epsilon}_{|2}]$, where $H_{|0}$ is given by Eq. \eqref{eq:H_0} but now evaluated in the new barred variables, and the quadratic contribution of the inhomogeneous modes is
\begin{subequations}\label{quadrahami}
\begin{align}
H^{\vec n,\epsilon}_{|2} &= \frac{e^{-\alpha}}{2}\big[E^n_{\bar\pi\bar\pi}\pi_{\bar f_{\vec n,\epsilon}}^2+2E^n_{\bar f \bar \pi}\bar f_{\vec n,\epsilon}\pi_{\bar f_{\vec n,\epsilon}}+E^n_{\bar f \bar f}\bar f_{\vec n,\epsilon}^2\big],\\
E^n_{\bar \pi \bar \pi} &= 1-\frac{3}{\omega_n^2}e^{-4\bar\alpha}\pi_{\bar\varphi}^2, \\
E^n_{\bar f \bar \pi} &= -\frac3{\omega_n^2}e^{-6{\bar\alpha}}\pi_{\bar\varphi}\big(e^{6\bar\alpha}\bar m^2\bar\varphi-2\pi_{\bar\alpha}\pi_{\bar\varphi}\big), \\
E^n_{\bar f \bar f} &= \omega_n^2+\bar m^2e^{2\bar\alpha}-\frac12e^{-4\bar\alpha}\big(\pi_{\bar\alpha}^2+15\pi_{\bar\varphi}^2+3e^{6\bar\alpha}\bar m^2\bar\varphi^2\big) -\frac3{\omega_n^2}e^{-8\bar\alpha}\big(e^{6\bar\alpha}\bar m^2\bar\varphi-2\pi_{\bar\alpha}\pi_{\bar\varphi}\big)^2.
\end{align}
\end{subequations}

To conclude this section, let us relate the canonical variables $(\bar f_{\vec n,\epsilon},\pi_{\bar f_{\vec n,\epsilon}})$ for the inhomogeneous modes  with the MS gauge invariants. In any gauge, the mode coefficients of the MS configuration field variable are \cite{mukhanov,hybridflat}
\begin{equation}\label{Mmode}
v_{\vec n,\epsilon} = e^\alpha\left[f_{\vec n,\epsilon}+\frac{\pi_{\varphi}}{\pi_\alpha}(a_{\vec n,\epsilon}+b_{\vec n,\epsilon})\right].
\end{equation}
Particularizing this expression to our longitudinal gauge, and introducing a conjugate momentum, we obtain the mode pairs
\begin{subequations}\label{eqs:MStrans}
\begin{align}
v_{\vec n,\epsilon} &= A_n\bar f_{\vec n,\epsilon}+B_n\pi_{\bar f_{\vec n,\epsilon}}, \\
\pi_{v_{\vec n,\epsilon}} &= C_n\bar f_{\vec n,\epsilon}+D_n\pi_{\bar f_{\vec n,\epsilon}},
\end{align}
\end{subequations}
where
\begin{subequations}\label{canonicoeff}
\begin{align}
A_n &= 1+\frac3{\omega_n^2}e^{-4\bar\alpha}\frac{\pi_{\bar\varphi}}{\pi_{\bar\alpha}}\big(e^{6\bar\alpha}\bar m^2\bar\varphi-2\pi_{\bar\alpha}\pi_{\bar\varphi}\big), \\
B_n &= \frac3{\omega_n^2}e^{-2\bar\alpha}\frac{\pi_{\bar\varphi}^2}{\pi_{\bar\alpha}}, \\
C_n &= -3e^{-2\bar\alpha}\frac{\pi_{\bar\varphi}^2}{\pi_{\bar\alpha}}-\frac3{\omega_n^2}e^{-6\bar\alpha}\frac1{\pi_{\bar\alpha}}\bigg[e^{12\bar\alpha}\bar m^4\bar\varphi^2+2\pi_{\bar\varphi}^2\big(2\pi_{\bar\alpha}^2-3\pi_{\bar\varphi}^2\big)\bigg] +\frac{3}{\omega_n^2} \bar m^2\bar\varphi\frac{\pi_{\bar\varphi}}{\pi_{\bar\alpha}^2}\big(4\pi_{\bar\alpha}^2-3\pi_{\bar\varphi}^2\big), \\
D_n &= 1-\frac3{\omega_n^2}e^{-4\bar\alpha}\frac{\pi_{\bar\varphi}}{\pi_{\bar\alpha}}\bigg[e^{6\bar\alpha}\bar m^2\bar\varphi-\frac{\pi_{\bar\varphi}}{\pi_{\bar\alpha}}\big(2\pi_{\bar\alpha}^2-3\pi_{\bar\varphi}^2)\bigg].
\end{align}
\end{subequations}
At this stage, a comment is in order. The expression of the MS momentum given here extends that given in Ref. \cite{hybridflat}, in the sense that both coincide only when the classical constraint $H$ is imposed, or, at the considered perturbative order, modulo the constraint $H_{|0}$ in the expression of the coefficients $C_n$ and $D_n$ as functions of the homogeneous variables.

The above relation between the MS pairs $(v_{\vec n,\epsilon},\pi_{v_{\vec n,\epsilon}})$ and the variables $(\bar f_{\vec n,\epsilon},\pi_{\bar f_{\vec n,\epsilon}})$ is a canonical transformation for fixed homogeneous variables. Actually, it is possible to prove that this transformation (with fixed homogeneous sector) can be implemented as a unitary one in the Fock representation selected by the choice of annihilation and creation-like variables that one would naturally construct from $(\bar f_{\vec n,\epsilon},\pi_{\bar f_{\vec n,\epsilon}})$ by disregarding the mass term of the scalar field. In the next section, we will extend this transformation to a canonical one not just on the inhomogeneities, but in the entire phase space of the reduced system.

\section{Formulation in terms of Mukhanov-Sasaki variables}\label{sec:MSvari}

The relation between the canonical pairs $(\bar f_{\vec n,\epsilon},\pi_{\bar f_{\vec n,\epsilon}})$ for the matter field Fourier coefficients and the MS pairs $(v_{\vec n,\epsilon},\pi_{v_{\vec n,\epsilon}})$ is canonical for fixed homogeneous variables, as we have commented, because it is easy to check that $A_nD_n-B_nC_n=1$ for all the possible values of $n$. Using this property, it is straightforward to obtain the inverse, given by
\begin{subequations}\label{invtransfo}
\begin{align}
\bar f_{\vec n,\epsilon} &= D_n v_{\vec n,\epsilon}- B_n\pi_{v_{\vec n,\epsilon}}, \\
\pi_{\bar f_{\vec n,\epsilon}} &= - C_n v_{\vec n,\epsilon}+A_n\pi_{v_{\vec n,\epsilon}}.
\end{align}
\end{subequations}
We will now complete this relation into a canonical transformation in the reduced phase space of the system, treated at quadratic perturbative order in the action.

Let us call $\{{\bar q}_A\}=\{\bar \alpha , \bar \varphi \}$, i.e., the barred homogeneous configuration variables, and $\bar \pi_{q_A}$ their canonical momenta. A simple calculation, using integration by parts, shows that, up to time integrals of total derivatives and neglecting cubic and higher contributions of the perturbations in the action, the Legendre term that contains the information about the symplectic structure can be rewritten:
\begin{equation} \label{symple}
\int  dt \Big[ \sum_A {\dot{\bar q}}_A \bar \pi_{q_A} + \sum_{\vec n, \epsilon}  {\dot{\bar f}}_{\vec n,\epsilon} \pi_{\bar f_{\vec n,\epsilon}} \Big]=  \int  dt \Big[ \sum_A {\dot{\tilde q}}_A \tilde \pi_{q_A} + \sum_{\vec n, \epsilon}  {\dot{v}}_{\vec n,\epsilon} \pi_{v_{\vec n,\epsilon}} \Big],
\end{equation}
where

\begin{align}\label{correcthomovari}\nonumber
{\tilde q}_A &= {\bar q}_A+ \frac{1}{2}\sum_{\vec n, \epsilon}  \bar f_{\vec n,\epsilon} \left(\partial_{\bar \pi_{q_A}}\pi_{\bar f_{\vec n,\epsilon}}\right)- \frac{1}{2}\sum_{\vec n, \epsilon}  \left(\partial_{\bar \pi_{q_A}}\bar f_{\vec n,\epsilon}\right) \pi_{\bar f_{\vec n,\epsilon}},\\
\tilde \pi_{q_A} &= \bar \pi_{q_A}- \frac{1}{2}\sum_{\vec n, \epsilon}  \bar f_{\vec n,\epsilon} \left(\partial_{{\bar q}_A}\pi_{\bar f_{\vec n,\epsilon}}\right) +\frac{1}{2} \sum_{\vec n, \epsilon}  \left(\partial_{{\bar q}_A}\bar f_{\vec n,\epsilon}\right)  \pi_{\bar f_{\vec n,\epsilon}}.
\end{align}
In these expressions, the partial derivatives are taken regarding $(\bar f_{\vec n,\epsilon},\pi_{\bar f_{\vec n,\epsilon}})$ as functions of the MS pairs, of $\bar q_A$, and of $\bar \pi_{q_A}$, as given by relations \eqref{invtransfo}.

From this result, it immediately follows that, at the perturbative order of our truncation, the set formed by the new homogeneous variables $(\tilde q_A, \tilde \pi_{q_A})=(\tilde \alpha , \tilde \varphi, \pi_{\tilde \alpha}, \pi_{\tilde \varphi})$ and the MS pairs $(v_{\vec n,\epsilon},\pi_{v_{\vec n,\epsilon}})$ is a canonical set in the phase space of the system. In other words, Eqs. \eqref{invtransfo} and \eqref{correcthomovari} are a canonical transformation in this phase space, at the relevant perturbative order.

In order to reformulate the system in terms of these new canonical variables, in which the inhomogeneities are described by gauge invariants, we still have to obtain the new expression of the only constraint remaining in the model, namely, the zero mode of the Hamiltonian constraint. For this, we first write the quadratic perturbative contribution $H^{\vec n,\epsilon}_{|2}$ as a function of the new variables, keeping just quadratic terms in the inhomogeneous modes. This can be easily done by: i) substituting in Eq. \eqref{quadrahami} the expression of the old variables $(\bar f_{\vec n,\epsilon},\pi_{\bar f_{\vec n,\epsilon}})$ in terms of the MS pairs [using Eq. \eqref{invtransfo}], and ii) replacing in the resulting expression the old homogeneous variables with the new ones, since their difference is quadratic in the inhomogeneities and is not significant at the considered perturbative order for $H^{\vec n,\epsilon}_{|2}$. In addition, we rewrite the other contribution to the constraint, $H_{|0}$, as a function of the new variables at the analyzed order in the inhomogeneous perturbations. Recalling that originally $H_{|0}$ was evaluated at the old homogeneous variables, and realizing that the difference of these variables with their new counterparts is quadratic in the MS modes, it is straightforward to conclude (e.g., by a series expansion of $H_{|0}$) that, at the mentioned truncation order, the desired contribution is provided by the evaluation of the homogeneous constraint $H_{|0}$ at the new homogeneous variables $(\tilde q_A, \tilde \pi_{q_A})$ plus a quadratic term in the perturbations given by the variation of $H_{|0}$ around those homogeneous variables multiplied by the variation of such variables produced by our change of canonical set. Combining these results, we get
\begin{subequations}\label{newH}
\begin{align}
H &= N_0\Big[H_{|0}({\tilde q}_A, \tilde \pi_{q_A})+{\tilde H}_{|2}({\tilde q}_A, \tilde \pi_{q_A},v_{\vec n,\epsilon},\pi_{v_{\vec n,\epsilon}})\big],\\
{\tilde H}_{|2} = \sum_{\vec n,\epsilon}{\tilde H}^{\vec n,\epsilon}_{|2} &= \sum_A \Big\{[{\bar q}_A - {\tilde q}_A]\partial_{\bar q_A} H_{|0}(\tilde q_A, \tilde \pi_{q_A}) +[{\bar \pi}_{q_A} - {\tilde \pi}_{q_A}]\partial_{\bar \pi_{q_A}} H_{|0}({\tilde q}_A, \tilde \pi_{q_A}) \Big\}\nonumber \\
&+ \sum_{\vec n,\epsilon}H^{\vec n,\epsilon}_{|2}({\tilde q}_A, \tilde \pi_{q_A}, \bar{f}_{\vec n,\epsilon}[{\tilde q}_A, \tilde \pi_{q_A},v_{\vec n,\epsilon},\pi_{v_{\vec n,\epsilon}}],\pi_{\bar f_{\vec n,\epsilon}}[{\tilde q}_A, \tilde \pi_{q_A},v_{\vec n,\epsilon},\pi_{v_{\vec n,\epsilon}}]),
\end{align}
\end{subequations}
with $(\bar{f}_{\vec n,\epsilon},\pi_{\bar f_{\vec n,\epsilon}})$ in the last formula given by Eq. \eqref{invtransfo} evaluated at ${\bar q}_A={\tilde q}_A$ and $\bar \pi_{q_A}=\tilde \pi_{q_A}$.

Alternatively, the expression for ${\tilde H}_{|2} $ can be obtained by considering our change of variables for the inhomogeneous modes as a time dependent canonical transformation for given homogeneous variables, whose time dependence is ruled in turn by the homogeneous contribution to the constraint $H_{|0}$. One can then apply the usual formulas for the change of Hamiltonian under canonical transformations which depend on time. The result is indeed the same that we have displayed above. This provides independent confirmation of the calculations and additional confidence in the consistency of our discussion.

A lengthy but direct computation leads then to the following formula for the quadratic contributions of the MS variables:
\begin{equation}\label{MSH2}
{\tilde H}^{\vec n,\epsilon}_{|2} = \frac{e^{-\tilde \alpha}}{2} \left\{\pi_{v_{\vec n,\epsilon}}^2 + \left[\omega_n^2+ e^{-4\tilde \alpha} \left( 19 \pi_{\tilde \varphi}^2 - 18\frac{\pi_{\tilde \varphi}^4}{\pi_{\tilde \alpha}^2} \right) + {\bar m}^2 e^{2\tilde \alpha} \left(1 -2 \tilde {\varphi}^2 -12 \tilde\varphi \frac{\pi_{\tilde \varphi} }{\pi_{\tilde \alpha}}\right)\right] v_{\vec n,\epsilon}^2\right\}.
\end{equation}
In arriving at this simple expression, we have used that $H_{|0}$ vanishes up to perturbative corrections. We notice that this quadratic Hamiltonian for the inhomogeneities contains no crossed term between the MS configuration variables and their momenta. Moreover, if one introduces unscaled MS variables $V_{\vec n,\epsilon}=e^{-\tilde \alpha}v_{\vec n,\epsilon}$ like those employed in the description of Refs. \cite{AAN1,AAN2}, with momenta given by $\pi_{V_{\vec n,\epsilon}}=e^{\tilde \alpha }\pi_{v_{\vec n,\epsilon}}+e^{-\tilde \alpha }\pi_{\tilde \alpha} v_{\vec n,\epsilon}$, and computes the corresponding Hamiltonian (either by considering this scaling as a time dependent canonical transformation of the inhomogeneous modes, or by completing it into a canonical transformation in the entire phase space of the system), one would obtain the same result as in Eqs. (2.5), (A3), and (A4) of the mentioned work \cite{AAN2} (taking into account the choice of lapse and homogeneous variables used there, and with the sum over discrete modes transformed into an integral for the case of noncompact flat topology). Thus, as expected, Eq. \eqref{MSH2} is just the counterpart of the MS Hamiltonian for the scaled inhomogeneous variables.

\section{Quantization}\label{sec:quantization}

In this section, we discuss the quantization of the symplectic manifold which describes our cosmological system, and of the Hamiltonian constraint to which it is subject. Physical states would be obtained as solutions to this constraint, imposed \`a la Dirac. In order to carry out this quantization, we combine loop and Fock techniques, according to our hybrid approach. The strategy is similar to that explained in Ref. \cite{hybridflat}; therefore, we only point out the essential steps. We first introduce a loop quantization of our homogeneous variables $(\tilde q_A, \tilde \pi_{q_A})$. For this, we adapt the parametrization of this homogeneous sector of the phase space to the standard one in LQC, in which the degrees of freedom of the geometry are described by an $su(2)$ connection and a densitized triad \cite{lqg}. In FRW cosmologies, these are respectively determined by two dynamical variables, $c$ and $p$, which are canonical in the sense that their Poisson bracket is equal to $8\pi G\gamma/3$, where $\gamma$ is the Immirzi parameter \cite{Immirzi}. Their relation with the variable $\tilde \alpha$ and its momentum in homogeneous and isotropic settings in the absence of inhomogeneous perturbations  ---which we extend to our situation as a definition of the variables that are to be quantized with the methods of LQC--- is
\begin{equation}\label{eq:homchange}
|p| = l_0^2\sigma^2e^{2\tilde\alpha},\quad pc = -\gamma l_0^3\sigma^2\pi_{\tilde \alpha}.
\end{equation}
The sign of $p$ determines the orientation of the triad, but we obviate it here because it will not play a relevant role in our quantization. In terms of this triad variable, we also introduce the homogeneous volume $V=| p|^{3/2}$, and the proportional variable
\begin{equation}\label{vp}
v=\sgn{(p)} \frac{|p|^{3/2}}{2\pi G \gamma  \sqrt{\Delta}},
\end{equation}
where $\sgn$ denotes the sign function and $\Delta$ is the minimum nonzero eigenvalue allowed for the area operator in LQG \cite{area}.

In addition, for the homogeneous variables related to the matter scalar field, we adopt the following scaling by a constant, in order to facilitate the comparison with the LQC literature:
\begin{equation}
\phi = \frac{\tilde \varphi}{l_0^{3/2}\sigma},\quad \pi_\phi = l_0^{3/2}\sigma\pi_{\tilde \varphi}.
\end{equation}

For the homogeneous degrees of freedom in the geometry, we introduce a quantization based on the so-called improved dynamics of LQC \cite{APS2} and on the quantization prescription of Ref. \cite{MMO} (usually called MMO prescription, after the initials of the authors Mart\'{i}n-Benito--Mena Marug\'an--Olmedo). This quantization is easy to specify in the $v$-representation in which the operator counterpart of the variable $v$ acts by multiplication. Defining as kinematical Hilbert space for the homogeneous sector of the geometry the Hilbert space ${\mathcal H}_{\mathrm{kin}}^{\mathrm{grav}}$ obtained by completing the span of all the eigenstates of $v$ (i.e., the set $\{ |v \rangle , v\in \mathbb{R}\}$) with the discrete norm $\langle v_1 | v_2 \rangle = \delta_{v_1,v_2}$, we introduce on it the operators with action
\begin{equation}
\hat v |v\rangle = v |v\rangle, \quad \hat N_{\bar\mu}|v\rangle = |v+1\rangle.
\end{equation}
For simplicity, we fix the reduced Planck constant $\hbar$ equal to the unit in all our discussion. The displacement operator $\hat N_{\bar\mu}$ provides the quantum representation of the nontrivial holonomy components along edges with fiducial length (with respect to the reference metric $^0h_{ij}$) equal to $l_0\bar \mu$, with $\bar \mu=\sqrt{\Delta/p}$, so that the physical area enclosed in a square formed by edges of this kind is precisely the gap area $\Delta$. It is easy to check that $b=\bar\mu c$ is canonically conjugate to the variable $v$ under Poisson brackets, with $\{b, v\}=2$. The displacement operator can be regarded as a representation of the holonomy component $\exp{(-ib/2)}$ in the improved dynamics formalism. On the other hand, for the homogeneous sector of the matter field, we adopt a standard representation with kinematical Hilbert space $\mathcal H_\mathrm{kin}^\mathrm{matt}$ given by the space $L^2(\mathbb{R},d\phi)$ of square integrable functions on the homogeneous field configuration with the Lebesgue metric, on which $\phi$ acts by multiplication and $\pi_{\phi}$ as $-i$ times the derivative with respect to $\phi$.

The contribution of the homogeneous degrees of freedom to the zero mode of the Hamiltonian constraint is represented by the operator \cite{MMO,MOP,hybridflat}:
\begin{equation}\label{eq:C_0}
\hat H_{|0} = \frac{\sigma}{2}\widehat{\left[\frac1{V}\right]}^{1/2}\hat{\mathcal C}_0\widehat{\left[\frac1{V}\right]}^{1/2}.
\end{equation}
The inverse-volume operator $\widehat{[1/V]}$ is the cube of the regularized operator
\begin{equation}
\widehat{\left[\frac1{V}\right]}^{1/3}=\widehat{\left[\frac1{\sqrt{|p|}}\right]} = \frac{3}{2 (2\pi\gamma G\sqrt\Delta)^{1/3}}\widehat{\sgn(v)}{|\hat v|}^{1/3}\big(\hat N_{-\bar\mu}{|\hat v|}^{1/3}\hat N_{\bar\mu}-\hat N_{\bar\mu}{|\hat v|}^{1/3}\hat N_{-\bar\mu}\big),
\end{equation}
which in fact commutes with the volume operator itself. Note that $\hat N_{-\bar\mu}$ is the inverse of $\hat N_{\bar\mu}$. On the other hand,
\begin{equation}
\label{eq:calC_0}
\hat{\mathcal C}_0 = {\hat \pi}_\phi^2- \hat{\mathcal H}_0^{(2)},
\end{equation}
where
\begin{subequations}
\begin{align}\label{eq:calH_0}
\hat{\mathcal H}_0^{(2)} &= \frac{3}{4\pi G \gamma^2}\hat\Omega_0^2-2 \hat V^2 W(\hat\phi); \quad W(\hat\phi)=\frac{1}{2}m^2\hat\phi^2,\\
\label{eq:Omega}
\hat\Omega_0 &= \frac1{4i\sqrt\Delta}\hat V^{1/2}\big[\widehat{\sgn(v)}\big(\hat N_{2\bar\mu}-\hat N_{-2\bar\mu}\big)+\big(\hat N_{2\bar\mu}-\hat N_{-2\bar\mu}\big)\widehat{\sgn(v)}\big]\hat V^{1/2}.
\end{align}
\end{subequations}
We have called $W(\phi)$ the potential of the scalar field, so that the discussion can be extended to situations beyond the mass contribution analyzed in detail here. On the other hand, the operator $\hat\Omega_0$ represents in this quantization the classical quantity $\Omega_0=pc$ once the latter has been approximated in terms of holonomies by $2\pi G \gamma v\sin{b}$. Its square, $\hat\Omega_0^2$, annihilates the zero-volume state $| v=0 \rangle$ and leaves invariant its orthogonal complement. Since the inverse-volume operator also annihilates that state, in the pure FRW sector of the system and as far as one is searching for solutions to the constraint, the analysis can be restricted to the mentioned orthogonal complement of $| v=0 \rangle$. Moreover, once this state is removed, one can establish a bijection between solutions to the constraint $\hat H_{|0}$ and solutions to $\hat {\mathcal C}_0$, which is much simpler to impose \cite{MMO}. Actually, the same procedure can be followed as well when the quadratic contributions of the inhomogeneities are introduced in the (zero mode of) the Hamiltonian constraint, because the action of this constraint again annihilates the zero-volume state, which decouples from its complement \cite{hybridflat}.

On the other hand, the action of $\hat\Omega_0^2$, and {\it a fortiori} that of $\hat{\mathcal C}_0$, superselects the kinematical Hilbert space of the homogeneous geometry sector. In fact, this action leaves invariant the subspaces $\mathcal H_\mathrm{\varepsilon}^\pm$ (which are separable, in contrast with the original $\mathcal H_{\mathrm{kin}}^{\mathrm{grav}}$) formed by states with support on the semilattices $\mathcal L_\mathrm{\varepsilon}^\pm=\{v=\pm(\varepsilon+4n)|n\in\mathbb N\}$, where $\varepsilon\in(0,4]$. Notice that, in each of these superselection sectors, the triad orientation does not change and the homogeneous volume $v$ has a strictly positive minimum (or negative maximum)\footnote{These properties are not shared by the prescription put forward in Refs. \cite{APS1,APS2}, whose superselection sectors are entire lattices.}. In the following, we will restrict the discussion, e.g., to semilattices with positive sign of $v$.

Let us consider now the representation of the quadratic contribution of the inhomogeneities to the zero mode of the Hamiltonian constraint. We first notice that, at the adopted truncation order and taking into account Eqs. \eqref{eq:calC_0} and \eqref{eq:calH_0}, we can substitute the value of $\pi_{\phi}^2$ in the expression of ${\tilde H}_{|2}$ with ${\mathcal H}_0^{(2)}=-2  V^2 W(\phi)+3\Omega_0^2/(4\pi G \gamma^2)$, represented quantum mechanically by $\hat{\mathcal H}_0^{(2)}$. This substitution will prove very convenient if one wants to use $\phi$ as an internal time in the system. The difference in the zero mode of the Hamiltonian constraint caused by this substitution is just of quartic order in the perturbations (because $\pi_{\phi}^2={\mathcal H}_0^{(2)}$ up to quadratic order terms). Hence, it can indeed be neglected. In fact, since $\pi_{\phi}^2$ is positive, we can go further and substitute ${\mathcal H}_0^{(2)}$ with its positive part, because it is only when this quantity is positive that the relation $\pi_{\phi}^2={\mathcal H}_0^{(2)}$ can be satisfied. We will call this positive part ${\mathcal H}_0^{2}$, and $\hat{\mathcal H}_0^{2}$ its operator representation, determined as the projection of $\hat{\mathcal H}_0^{(2)}$ in the positive part of its spectrum. We assume that this operator $\hat{\mathcal H}_0^2$ can be defined (generally in a non-unique way) as self-adjoint in ${\mathcal H}_\mathrm{kin}^\mathrm{grav}$ for every value of $\phi$, as has been argued in the literature \cite{lqc2}\footnote{An alternate prescription, which we will find specially interesting for the factor ordering discussed in Sec. \ref{orderin}, is to identify instead $\hat{\mathcal H}_0$ with the (self-adjoint) operator that dictates the evolution in $\phi$ of the positive frequency solutions of the homogeneous model obtained by group averaging \cite{AAN2,Marolf}.}. After this procedure, ${\tilde H}_{|2}$ becomes a linear function of the momentum $\pi_{\phi}$, of the generic form
\begin{equation}\label{quadhamMS}
{\tilde H}_{|2}\equiv \frac{\sigma}{2 V}\sum_{\vec n,\epsilon}{\mathcal C}^{\vec n,\epsilon}_{2}; \quad  {\mathcal C}^{\vec n,\epsilon}_{2}= -\Theta^{\vec n,\epsilon}_{\mathrm e}- \Theta^{\vec n,\epsilon}_{\mathrm o}\pi_{\phi}.
\end{equation}
In our case, we obtain
\begin{subequations}\label{thetaope}
\begin{align}
\frac{1}{V^{2/3}} \Theta^{\vec n,\epsilon}_{\mathrm e}&= - \left\{\frac{4\pi G}{3V^{4/3}} {\mathcal H}_0^2 \left(19-24\pi G \gamma^2\frac{{\mathcal H}_0^2}{\Omega_0^2}\right)+V^{2/3}\left[W^{\prime\prime}(\phi)-\frac{16 \pi G}{3} W(\phi)\right]\right\} {\tilde v}^2_{\vec n,\epsilon}
\nonumber\\ &
- {\tilde \omega}_n^2{\tilde v}^2_{\vec n,\epsilon}-\pi^2_{{\tilde v}_{\vec n,\epsilon}},\\ \label{thetaope2}
\frac{1}{V^{2/3}} \Theta^{\vec n,\epsilon}_{\mathrm o}&=- 16 \pi G \gamma V^{2/3}  \frac{W^{\prime}(\phi)}{\Omega_0}{\tilde v}^2_{\vec n,\epsilon},
\end{align}
\end{subequations}
where the prime denotes the derivative with respect to $\phi$ in the potential $W$, we have defined ${\tilde \omega}_n=l_0 \omega_n$, and we have rescaled the MS variables by a constant number, namely:
\begin{equation}\label{scaleMS}
{\tilde v}_{\vec n,\epsilon}=\frac{v_{\vec n,\epsilon}}{\sqrt{l_0}},\quad \pi_{\tilde v_{\vec n,\epsilon}}=\sqrt{l_0}\pi_{v_{\vec n,\epsilon}}.
\end{equation}

For the factors in this contribution that depend on the homogeneous variables, and which are affected in principle by some quantization ambiguities, we will introduce a symmetric factor ordering that tries to respect, as far as possible, the assignations of representation made in the FRW part of the system. Specifically, we adopt the prescriptions explained in Ref. \cite{hybridflat}: i) for products $f(\phi)\pi_{\phi}$, where $f$ is an arbitrary function, we adopt a symmetric factor ordering of the form $\{f(\hat \phi) \hat\pi_\phi+\hat\pi_\phi f(\hat\phi)\}/2$; ii) for factors of the homogeneous volume, we adopt an algebraic symmetrization, so that terms like $V^rg(cp)$, where $g$ is any function and $r$ a real number, is promoted to the operator $\hat V^{r/2}\hat g\hat V^{r/2}$; besides, this algebraic symmetric factor ordering is also taken for powers of the inverse volume; iii) for even powers of the phase space variable $\Omega_0=cp$, we represent this quantity by the same powers of the operator $\hat \Omega_0$, as in FRW; and, finally, iv) for odd powers of $\Omega_0=cp$, let us say $\Omega_0^{2k+1}$ with $k$ equal to an integer, we choose the representation $|\hat\Omega_0|^k \hat\Lambda_0|\hat\Omega_0|^k$, where $|\hat\Omega_0|$ is the positive operator provided by the square root of $\hat\Omega_0^2$ and
\begin{equation}
\hat\Lambda_0 = \frac1{8i\sqrt\Delta}\hat V^{1/2}\big[\widehat{\sgn(v)}\big(\hat N_{4\bar\mu}-\hat N_{-4\bar\mu}\big)+\big(\hat N_{4\bar\mu}-\hat N_{-4\bar\mu}\big)\widehat{\sgn(v)}\big]\hat V^{1/2}.
\end{equation}
Note that this operator is defined in a similar way as $\hat \Omega_0$, but with holonomies of double fiducial length. As a result, the displacements in $v$ that its action may cause are always multiples of four units, so that it leaves invariant the superselection semilattices $\mathcal L_\mathrm{\varepsilon}^\pm$ of the homogeneous geometry. Had we just replaced $\hat\Lambda_0$ with $\hat \Omega_0$, without doubling the fiducial length of the holonomy edges, the displacements might have been of only two units, and hence the superselections sectors of FRW would not have been respected. Actually, our strategy parallels the usual choice made in the LQC description of FRW universes when one represents the Hubble parameter \cite{MOP}.

Following the hybrid approach, we adopt a Fock representation for the inhomogeneous modes, in a quantization that is selected by the criteria of: i) vacuum invariance under the spatial isometries, and ii) unitary implementability of the dynamical evolution in the regime in which one recovers a QFT in a curved background (in any finite time interval) \cite{uniqueness1,uniqueness2}. As we have mentioned, these criteria pick out the canonical pairs of variables that we have chosen for the descriptMarolfion of the inhomogeneous perturbations \cite{uniqueness1} ---obviously up to a constant scaling of all the configuration variables and the opposite scaling of their momenta. Any other choice of canonical pairs among those related with ours by a scaling of the scalar field using a function of the homogeneous variables (which might even be explicitly time dependent) would simply not allow for a unitary dynamics in the mentioned QFT regime, regardless of the complex structure chosen to construct the Fock representation (this is the case, for instance, of the canonical pairs chosen in Refs. \cite{AAN1,AAN2}). Although one may always renounce to unitarity, this would imply that the Heisenberg description of the inhomogeneities would be inequivalent to a Schr\"odinger description. On the contrary, with our criteria we do not only remove the ambiguity in splitting the dependence of the field modes on the homogeneous and inhomogeneous variables, but we assure a unitary implementability of the evolution and a standard quantum mechanical interpretation in the sector where a QFT in a (generally effective) background is recovered. Besides, with our choice of canonical pairs for the inhomogeneous modes, our invariance and unitarity criteria select a family of Fock representations that are all unitarily equivalent \cite{uniqueness2}. This family contains the representation in which the annihilation and creation-like variables for the modes are those naturally associated with harmonic oscillators of frequency $\tilde \omega_n$, namely,
\begin{equation}\label{annihi}
a_{{\tilde v}_{\vec n,\epsilon}}=\frac{1}{\sqrt{2\tilde \omega_n}}(\tilde \omega_n {\tilde v}_{\vec n,\epsilon}+i\pi_{{\tilde v}_{\vec n,\epsilon}})
\end{equation}
and their complex conjugates as creation-like variables.

Any representation invariant under the spatial isometries and in the class of (unitary) equivalence of the one determined by the above annihilation and creation-like variables is acceptable. Although they are all (unitarily) equivalent as far as the representation of functions of the field in the Weyl algebra is concerned (i.e., exponentials of linear combinations of the field and its momentum and, given the continuity of the representation, those linear combinations themselves), the definition of other field operators may depend on the particular representation taken in the selected class, as it is the case of quadratic operators like the one corresponding to the contribution of the inhomogeneities in the zero mode of the Hamiltonian constraint. Conditions on physically relevant operators, like e.g. this Hamiltonian, may remove the still existing freedom in the choice of Fock representation, at least partially. Natural conditions are that the considered operators are well defined and essentially self-adjoint. Other properties concerning their regularization may be important, although our viewpoint is that the regularization schemes should arise directly from the quantization of the system, and not as techniques imported from QFT in curved backgrounds, as it is usually conceived that such techniques should find their justification in a more fundamental quantum theory of spacetime, and LQC is assumed to be a framework of that kind, at least to some extent.

Let us then suppose that (either by imposing additional conditions on physical operators or by mere choice) we take a Fock quantization in the above class of representations that are invariant under the spatial isometries, and in this way, in particular, we promote to operators the variables ${\tilde v}^2_{\vec n,\epsilon}$ and $\pi^2_{{\tilde v}_{\vec n,\epsilon}}$ appearing in Eq. \eqref{thetaope}. Let us also call ${\mathcal F}$ the corresponding Fock space. A basis for the space is formed by the occupancy-number states, $|{\mathcal N}\rangle$, in which a finite number of modes presents a kind of {\it particle} excitation as  interpreted in terms of the natural annihilation and creation operators of the representation \cite{hybridflat}. The total kinematical Hilbert space of our quantization is simply the product of those of the homogeneous and inhomogeneous variables, $\mathcal H_\mathrm{kin}^\mathrm{tot} = \mathcal H_\mathrm{kin}^\mathrm{grav}\otimes\mathcal H_\mathrm{kin}^\mathrm{matt}\otimes\mathcal F$. Clearly, the zero mode of the Hamiltonian constraint has a nontrivial action on this space, since it does not respect its product structure, because the part that is quadratic in the perturbations mixes the homogeneous and the inhomogeneous sectors.
According to our discussion, this constraint can be written in the form $\hat {\mathcal C}=\hat {\mathcal C}_0+\sum_{\vec n,\epsilon}{\hat{\mathcal C}}^{\vec n,\epsilon}_2$, where the operators representing ${\mathcal C}^{\vec n,\epsilon}_2$,
\begin{equation}
{\hat{\mathcal C}}^{\vec n,\epsilon}_2= -\hat{\Theta}^{\vec n,\epsilon}_{\mathrm e}- \big(\hat{\Theta}^{\vec n,\epsilon}_{\mathrm o}{\hat \pi}_{\phi}\big)_S,
\end{equation}
are constructed with the prescriptions that we have explained above. The symbol $(\;)_S$ denotes symmetrization in the product of operators. This takes care of the product of ${\hat \pi}_{\phi}$ with functions of $\phi$: here, specifically, with the factor $W^ {\prime}(\phi)$ in Eq. \eqref{thetaope2}, if the potential of the field has a nonvanishing derivative. For later convenience, we also introduce the notation
\begin{subequations}
\begin{align}\label{C2MS}
\hat{\mathcal C}_2&=\sum_{\vec n,\epsilon}\hat{\mathcal C}^{\vec n,\epsilon}_2=-\hat{\Theta}_{\mathrm e}- \big(\hat{\Theta}_{\mathrm o}{\hat \pi}_{\phi}\big)_S,\\
\label{thetaopeMS}
\hat{\Theta}_{\mathrm e} &=\sum_{\vec n,\epsilon}\hat{\Theta}^{\vec n,\epsilon}_{\mathrm e},\quad \hat{\Theta}_{\mathrm o}
=\sum_{\vec n,\epsilon}\hat{\Theta}^{\vec n,\epsilon}_{\mathrm o}.
\end{align}
\end{subequations}

\section{Born-Oppenheimer ansatz}\label{BOa}

In this section, we will analyze the behavior of the possible physical states of the system whose dependence on the homogeneous degrees of freedom of the FRW geometry, on the one hand, and on the inhomogeneous modes, on the other hand, can be separated. This separation will be possible, essentially, because the two mentioned kinds of degrees of freedom will present different rates of variation with respect to the homogeneous part $\phi$ of the matter scalar field, regarded as an internal time for the system (at least in some intervals of the evolution). In this sense, we will say that we introduce an ansatz of BO type for the states. Specifically, we consider states with wave functions $\Psi$ of the form
\begin{equation}\label{BOans}
\Psi=\chi(V,\phi) \psi({\mathcal N},\phi),
\end{equation}
where the dependence on the MS variables has been included in terms of the label ${\mathcal N}$ of the basis of occupation-number states for the inhomogeneous modes. We note the dependence on $\phi$ of the two factors in the wave function.

Moreover, we assume that the part of the state that contains the dependence on the FRW geometry is determined by a state $\chi_0(V)$  of the homogeneous gravitational degrees of freedom at a fixed value $\phi_0$ of $\phi$, evolved with $\hat{\cal H}_0$ to other values of the homogeneous variable of the scalar field. More precisely, we only consider states $\chi_0(V)$ on which $\hat{\cal H}_0^{(2)}$ acts as its positive part; then $\hat{\cal H}_0$ can be defined as in the previous section and, at least when its variation with respect to $\phi$ is negligible, interpreted as the Hamiltonian for positive frequency states in the loop quantization of FRW after the deparametrization of the system, adopting $\phi$ as internal time\footnote{A similar procedure can be adopted for group averaging strategies, regarding the corresponding positive frequency Hamiltonian $\hat{\cal H}_0$ as a specific quantization prescription for FRW, affecting also the representation of the homogeneous contribution to the constraint.}. In summary,
\begin{equation}\label{chievolu}
\chi(V,\phi)={\mathcal P}\left[\exp{\left(i\int^{\phi}_{\phi_0} d\tilde{\phi}\,\hat{\mathcal H}_0( \tilde \phi)\right)}\right] \chi_0(V).
\end{equation}
The state $\chi_0$ is normalized to the unit in the inner product of the kinematical Hilbert space for the FRW geometry, ${\mathcal H}_\mathrm{kin}^\mathrm{grav}$. The symbol ${\mathcal P}$ denotes {\it time} ordering with respect to $\phi$, ordering that is needed in the definition of the exponentiated integral because $\hat{\mathcal H}_0$ generically depends on $\phi$ through the matter field potential. Notice that, provided that $\hat{\mathcal H}_0$ is self-adjoint for each value of $\phi$ as we have argued, the evolution that it generates is unitary. In addition, and although not strictly necessary for most of our following discussion, we will suppose that the state $\chi_0$ of the FRW geometry is so peaked that the corresponding state $\chi$ remains peaked for all considered values of $\phi$, and that its peak can be described with the equations of effective LQC for homogeneous and isotropic universes deduced for states with a considerable semiclassical behavior at very large volumes \cite{taveras}.

Let us then plug this ansatz in the constraint equation\footnote{Since solutions are not expected to belong to the kinematical Hilbert space, one should rather impose the constraint in the form $(\psi| \hat {\mathcal C}^{\dagger}=0$ on some kind of generalized states $(\psi|$, where the dagger denotes the adjoint. With this caveat, we continue our discussion without introducing adjoints, something that would complicate the notation even more.} $\hat {\mathcal C}\Psi=0$. If we {\emph{disregard possible nondiagonal elements in the homogeneous geometry variables}} (i.e, possible quantum transitions from $\chi$ to another state mediated by the action of the constraint), and consider only the diagonal part, that can be extracted by taking the inner product with the state $\chi$ in ${\mathcal H}_\mathrm{kin}^\mathrm{grav}$, we arrive at the result
\begin{equation}\label{constrainBO}
-\partial_{\phi}^2 \psi - i \left(2 \langle \hat{\mathcal H}_0 \rangle_{\chi} - \langle \hat{\Theta}_{\mathrm o} \rangle_{\chi}\right) \partial_{\phi}\psi=\left[\langle \hat{\Theta}_{\mathrm e}+ \big(\hat{\Theta}_{\mathrm o} \hat{\mathcal H}_{0}\big)_S\rangle_{\chi}
+i \langle   {\mathrm d}_{\phi}\hat{\mathcal H}_{0} - \frac{1}{2}{\mathrm d}_{\phi}\hat{\Theta}_{\mathrm o} \rangle_{\chi} \right] \psi.
\end{equation}
Here, $\langle \, \rangle_{\chi}$ is the expectation value on $\chi$, with respect to the inner product in ${\mathcal H}_\mathrm{kin}^\mathrm{grav}$, and ${\mathrm d}_{\phi}$ stands for what in the Heisenberg picture is the total derivative of an operator with respect to $\phi$; namely, for any operator $\hat{O}$, we have\footnote{We choose the sign of the evolution generator in accordance with our positive frequency convention.}
\begin{equation}\label{totalder}
{\mathrm d}_{\phi}  \hat{O} =    \partial_{\phi} \hat{O} -i [ \hat{\mathcal H}_{0},\hat{O}] .
\end{equation}
Notice that, in the case of $\hat{\mathcal H}_{0}$, the last term does not contribute because the commutator vanishes.

We see that this constraint equation would lead to a Schr\"odinger equation for the evolution of the inhomogeneities in $\phi$ provided that the following conditions are satisfied. a) $\langle \hat{\Theta}_{\mathrm o} \rangle_{\chi}$ has to be negligible as compared to $\langle \hat{\mathcal H}_0 \rangle_{\chi}$ in the term proportional to the derivative of $\psi$. In our perturbative approximation, this is always the case, if we insist on regarding the approximation as an asymptotic expansion (in the limit where a certain perturbative parameter vanishes), in which ${\mathcal H}_0$ is of the order of the unity. In practice, nonetheless, the approximation is acceptable if it is true that the quadratic contribution of the inhomogeneities given by $\hat{\Theta}_{\mathrm o}$ remains small when compared to the generator of the $\phi$-evolution in the FRW case ---additional comments can be found in Sec. \ref{orderin}. b) It may be possible to neglect the second derivative of $\psi$ in the  equation. This may be checked by self-consistency, because if one assumes that this happens, together with condition a), one can obtain the value of $\partial_{\phi}\psi$ from Eq. \eqref{constrainBO}. Deriving this value with respect to $\phi$, one can see whether the second derivative of the wave function of the perturbations is indeed negligible compared to the first derivative. We will return to this issue later in this section. In addition to all this, if the evolution of the inhomogeneities in $\phi$ is to be ruled by a real Hamiltonian (something necessary if we want it to become self-adjoint in the Fock space), one needs: c) The total $\phi$-derivative of $(2\hat{\mathcal H}_{0} - \hat{\Theta}_{\mathrm o})$ must be negligible compared to the contribution of the MS Hamiltonian.

If the three conditions were satisfied, we would get the Schr\"odinger equation
\begin{equation}\label{schroMS}
-i \partial_{\phi}\psi=\frac{\langle \hat{\Theta}_{\mathrm e}+ \big(\hat{\Theta}_{\mathrm o} \hat{\mathcal H}_{0}\big)_S\rangle_{\chi}}{2 \langle \hat{\mathcal H}_0 \rangle_{\chi}}\psi.
\end{equation}
Note that $\langle \hat{\mathcal H}_0 \rangle_{\chi}$ is just a function of $\phi$, and hence we can divide by it, if it is different from zero. The term in the right-hand side acting on $\psi$ can be interpreted in this approximation as the Hamiltonian that generates the dynamics of the perturbations in the internal time $\phi$. This Hamiltonian is just the MS Hamiltonian, with its dependence on the homogeneous geometry variables evaluated at the expectation values corresponding to the quantum state $\chi$, and divided by the expectation value of $\hat{\mathcal H}_0$. This last factor (as we will show below) can be seen as providing the change of time to $\phi$ in the peak trajectory of $\chi$.

Apart from differences in the Fock quantization and in the prescriptions used to define the quantum operators that appear in it, this Schr\"odinger equation resembles remarkably the evolution equation put forward for the perturbations in the dressed metric approach. The main discrepancy, in practice, is the range of validity deduced for it in the hybrid approach, summarized in conditions a)-c).

Returning to our previous discussion, suppose that we only admit the validity of condition a), something which, as we have explained, can always be justified on the basis of the perturbative hierarchy. We would then get
\begin{equation}\label{constrainBO2}
- i  \partial_{\phi}\psi=\frac{1}{2 \langle \hat{\mathcal H}_0 \rangle_{\chi}} \left[\langle \hat{\Theta}_{\mathrm e}+ \big(\hat{\Theta}_{\mathrm o} \hat{\mathcal H}_{0}\big)_S\rangle_{\chi}
+i \langle   {\mathrm d}_{\phi}\hat{\mathcal H}_{0} - \frac{1}{2}{\mathrm d}_{\phi}\hat{\Theta}_{\mathrm o} \rangle_{\chi} \right] \psi + \frac{1}{2 \langle \hat{\mathcal H}_0 \rangle_{\chi}} \partial_{\phi}^2 \psi ,
\end{equation}
and, deriving this expression with respect to $\phi$ and eliminating terms which are negligible perturbatively,
\begin{eqnarray}\label{constrainBO3}
&&\left[\frac{3\langle {\mathrm d}_{\phi}\hat{\mathcal H}_0 \rangle_{\chi}}{2\langle \hat{\mathcal H}_0 \rangle_{\chi}} -2 i \langle \hat{\mathcal H}_0 \rangle_{\chi}\right]\partial_{\phi}^2\psi =-\frac{\langle {\mathrm d}_{\phi}\hat{\mathcal H}_0 \rangle_{\chi}}{\langle \hat{\mathcal H}_0 \rangle_{\chi}} \left[2\langle \hat{\Theta}_{\mathrm e}+ \big(\hat{\Theta}_{\mathrm o} \hat{\mathcal H}_{0}\big)_S\rangle_{\chi}
+\frac{i}{2} \langle   3{\mathrm d}_{\phi}\hat{\mathcal H}_{0} - 2{\mathrm d}_{\phi}\hat{\Theta}_{\mathrm o} \rangle_{\chi}\right] \psi \nonumber\\
&&\quad+
\left[\langle {\mathrm d}_{\phi}\hat{\Theta}_{\mathrm e}+ {\mathrm d}_{\phi}\big(\hat{\Theta}_{\mathrm o} \hat{\mathcal H}_{0}\big)_S\rangle_{\chi}
+i \langle   {\mathrm d}_{\phi}^2\hat{\mathcal H}_{0} - \frac{1}{2}{\mathrm d}_{\phi}^2\hat{\Theta}_{\mathrm o} \rangle_{\chi} \right] \psi  + \partial_{\phi}^3 \psi .
\end{eqnarray}
With this equation, it is possible to see whether it is consistent to assume that each new derivative of $\psi$ with respect to $\phi$ is negligible compared to the previous one, and hence if condition b) is indeed satisfied.

It is easy to convince oneself, from the above analysis, that the validity of conditions b) and c) depends on how negligible the {\it total} derivatives of the operators $\hat{\mathcal H}_{0}$, $\hat{\Theta}_{\mathrm e}$, and $\hat{\Theta}_{\mathrm o}$ [and also $(\hat{\Theta}_{\mathrm o}\hat{\mathcal H}_{0})_S$] with respect to $\phi$ are. More precisely, a careful consideration of conditions b) and c), once the validity of condition a) has been accepted, indicates that one needs that the derivatives of the involved operators to be negligible compared to the MS Hamiltonian, in expectation values on $\chi$. Actually, one can relax condition c) and keep the contribution of $\langle {\mathrm d}_{\phi}\hat{\mathcal H}_{0}\rangle_{\chi}$ in Eq. \eqref{constrainBO2}, which may later be absorbed by a $\phi$-dependent change of norm in $\chi$. In that case, one can show that, rather than the mentioned derivative contribution, it is its square and $\langle {\mathrm d}_{\phi}^2\hat{\mathcal H}_{0}\rangle_{\chi}$ what has to be negligible compared with the expectation value of the MS Hamiltonian.

Nonetheless, before deciding to go on and carry out a detailed analysis of the circumstances under which the considered derivatives can be ignored in our equations, let us recall that these total derivatives contain two types of terms [see Eq. \eqref{totalder}]. One of them is a derivative with respect to the explicit dependence on $\phi$ of these operators. This dependence comes exclusively from the potential of the matter scalar field. If the derivatives of this potential are sufficiently small in the possible range of variation of $\phi$, all terms of this kind might be negligible at the desired order. For instance, if the potential is a mass term, phenomenologically the possible values for the mass are considerably small, and the derivatives of the potential might be treated as perturbative terms, e.g. by expressing the mass value as a certain power of the amplitude parameter of the inhomogeneous perturbations. But there is still a second type of terms in the analyzed derivatives, namely, the commutator of the operator with $\hat{\mathcal H}_{0}$. This commutator gives a nonvanishing contribution in the derivatives of the theta-operators appearing in the MS Hamiltonian. Since the dependence on the homogeneous variables of these theta-operators and of $\hat{\mathcal H}_{0}$ is only through the FRW geometry variables and $\phi$, the commutator in question gets nontrivial values only because of the contributions of the homogeneous geometry. Hence, the commutator can get relevant terms from the operator dependence of $\hat{\mathcal H}_{0}$ on $\hat \Omega_0^2$ and of the theta-operators on $\hat V$, and viceversa. Recall that the commutator of $\hat \Omega_0^2$ and $\hat V$ gives a term proportional to $\sin{(2b)}$ in the effective regime of LQC for FRW geometries\footnote{This term is also proportional to the homogeneous volume, but this additional factor may be compensated by a decrease in powers of this variable caused by the derivative with respect to $V$ taken in one of the operators that formed the commutator.}, a term which can be of order of the unit in some stages of the evolution. Actually, the Big Bounce would correspond to values of $\sin{b}$ equal to one, and would be preceded and followed by regions where the sine of $2b$ would be close to the unit value. It is precisely in those regions where different authors, studying the closure of the modified algebra of constraints in LQG and its consequences for cosmological perturbations, have claimed that the spacetime structure suffers from a change of signature \cite{cai,bojogianlufren,signature}. Independently of the possibility of this process of signature change, we see that there exist reasons to admit that these contributions to the commutators, and hence to the equations of the cosmological perturbations, may not be always negligible. Therefore, conditions b)-c) should be checked to confirm that they hold before one can approximate the evolution equation \eqref{constrainBO2} in our hybrid quantization by its Schr\"odinger version \eqref{schroMS}.

\section{Alternate factor ordering}\label{orderin}

In the preceding section, we have seen that, once the BO ansatz is introduced in the hybrid quantization, some terms in the constraint equation that must be neglected in order to arrive at a Schr\"odinger equation come from total derivatives of operators with respect to the internal time $\phi$. A second thought about these terms reveals that they arise in fact from factor ordering ambiguities in the quantization procedure. In other words, they can be absorbed by adopting a different factor ordering. Actually, the part of the total derivatives that is given by a commutator with the homogeneous Hamiltonian $\hat{\mathcal H}_{0}$ is clearly a quantum correction (which can be removed if one changes the order of the operators in the expression).
But something similar happens also with the partial derivatives of the operators with respect to $\phi$ in the expectation values over the homogeneous geometry: in the quantization that we discussed in the previous sections, these partial derivatives can be identified with the commutators of the considered operators with the momentum of the homogeneous part of the scalar field, $\hat \pi_{\phi}$. In the light of these comments, it seems natural to search for a different factor ordering in this quantization from which one can derive an evolution equation for the perturbations similar to that of the dressed metric approach \cite{AAN1,AAN2}. Recall, in this sense, that except for a different scaling of the inhomogeneous modes in the matter field and the associated MS variables, the quadratic contribution to the constraint $\Theta_{\mathrm e} + \Theta_{\mathrm o} \pi_{\phi}$ is just the MS Hamiltonian for the inhomogeneities which generates their evolution in the time $T$ with $dt=2VdT$ in the classical theory, with $t$ being the proper time [see Eq. \eqref{quadhamMS} and the definition of the homogeneous part of the lapse function in Eq. \eqref{lapse}].

As we have seen, the zero mode of the Hamiltonian constraint [up to a factor $\sigma/(2V)$] is given classically by ${\mathcal C}=\pi_{\phi}^2-{\mathcal H}_0^2-\Theta_{\mathrm e} - \Theta_{\mathrm o} \pi_{\phi}$ (where we have used an obvious notation for the classical phase space functions appearing in the constraint). It is straightforward to see that, at the considered truncation order, quadratic in the inhomogeneous modes, we have
\begin{equation}\label{factorize}
{\mathcal C}=\left[\pi_{\phi}+{\mathcal H}_0+\frac{1}{2}\left(\Theta_{\mathrm e} + \Theta_{\mathrm o}\pi_{\phi}\right) {\mathcal H}_0^{-1} \right]\left[\pi_{\phi}-{\mathcal H}_0 - \frac{1}{2}{\mathcal H}_0^{-1}\left(\Theta_{\mathrm e}+ \Theta_{\mathrm o} \pi_{\phi}\right)\right].
\end{equation}
If we regard our perturbative approximation as an asymptotic expansion, the terms of the form ${\mathcal H}_0^{-1}\left(\Theta_{\mathrm e}+ \Theta_{\mathrm o} \pi_{\phi}\right)$ can still be treated perturbatively as quadratic corrections. In practice, nonetheless, the results of the analysis will be meaningful if these terms are in fact small. This means that the product of the inverse of ${\mathcal H}_0$ by our original MS contributions must be small. This may involve complications in the sector of small values of ${\mathcal H}_0$ (quantum mechanically, in the region of the spectrum of the operator $\hat{\mathcal H}_0$ close to its kernel). We note that this sector has small values of the momentum of the homogeneous scalar field when the inhomogeneities are also small. This may be problematic for the numerical accuracy of the approximation with the alternate factor ordering that we are trying to adopt now.

It is also worth commenting that this situation is different from what we found in Sec. \ref{BOa}. There, we needed condition a) in order to deduce Eq. \eqref{constrainBO2}, but that condition required only that (in expectation values) the MS contribution $\Theta_{\rm o}$ be negligible compared to the homogeneous Hamiltonian ${\mathcal H}_0$. As one can check in Eq. \eqref{thetaope2}, $\Theta_{\rm o}$ is proportional to the derivative of the potential of the scalar field, which can be considerably small. In the studied case of a massive field, this derivative is $m^2\phi$. If one then takes into account that, in effective LQC for FRW universes, the absolute value of the homogeneous field is bounded from above for this potential by a number of the order of $1/m$ (see Ref. \cite{ASloan}), one concludes that, in the allowed range of variation, the derivative of the potential is at most of the order of the mass. In total, $\Theta_{\rm o}$ is a quadratic contribution in the inhomogeneous perturbations multiplied, in addition, by a factor of order $m$, leading to a really small quantity and justifying the validity of the commented condition a).

We can now quantize the constraint with the ordering of Eq. \eqref{factorize}, adopting for each factor, e.g., the prescriptions of previous sections. This factor ordering, though not symmetric, is specially appropriate if we are only interested in {\emph{perturbative}} solutions of positive frequency with respect to the variable $\phi$. For this type of positive $\phi$-frequency solutions, which must remain meaningful in the asymptotic limit of vanishing perturbations, the first factor (on the left) in the constraint equation cannot annihilate the quantum state. Its corresponding positive $\phi$-frequency solutions would be annihilated by $\hat \pi_{\phi}$ ---and hence belong to the kernel of $\hat {\mathcal H}_0$--- in the limit of a purely homogeneous truncation of the system. Remarkably, it is in the neighborhood of  this kernel where we pointed out the possibility that there existed practical problems with the perturbative approximation in the factor ordering considered here. With this caveat, the perturbative solutions $\Psi$ of positive $\phi$-frequency are determined as solutions of the equation
\begin{equation}\label{AANconstrain}
\left[\hat \pi_{\phi}-\hat {\mathcal H}_0 - \frac{1}{2}\hat {\mathcal H}_0^{-1/2}\left(\hat \Theta_{\mathrm e}+ \big(\hat \Theta_{\mathrm o} \hat\pi_{\phi}\big)_S\right)\hat {\mathcal H}_0^{-1/2}\right]\Psi=0.
\end{equation}
Note that we have adopted an algebraic symmetric factor ordering for the product of the operator $\hat {\mathcal H}_0^{-1} $ with the MS Hamiltonian, rather than other symmetrizations, so that we do not have to change the prescription for the representation of this MS Hamiltonian.

If we now introduce the BO ansatz \eqref{BOans} and \eqref{chievolu}, and {\emph{ignore nondiagonal elements} in the homogeneous geometry, considering only the diagonal part by taking the inner product in ${\mathcal H}_\mathrm{kin}^\mathrm{grav}$ with $\chi$, we arrive at the following evolution equation for the perturbations:
\begin{equation}\label{AANconstrain2}
-i\partial_{\phi}\psi= \frac{1}{2}\big\langle \hat {\mathcal H}_0^{-1/2}\Big(\hat \Theta_{\mathrm e}+ \big(\hat \Theta_{\mathrm o} \hat {\mathcal H}_0\big)_S\Big)\hat {\mathcal H}_0^{-1/2}-\frac{i}{2}\hat {\mathcal H}_0^{-1/2}{\mathrm d}_{\phi} \big(\hat \Theta_{\mathrm o}\hat {\mathcal
H}_0^{-1}\big)\hat {\mathcal H}_0^{1/2}\big\rangle_{\chi} \psi .
\end{equation}
This Schr\"odinger equation is similar to the evolution equation for the perturbations of the dressed metric approach. The differences with respect to the discussion in Refs. \cite{AAN1,AAN2} affect only the scaling of the inhomogeneous modes and the prescriptions for the quantization of the Hamiltonian in the right-hand side. In particular, the contribution of the derivative ${\mathrm d}_{\phi} (\hat \Theta_{\mathrm o} \hat {\mathcal H}_0^{-1})$ can be removed with a different choice of operator representation for the product of $\Theta_{\mathrm o}$, ${\mathcal H}_0^{-1}$, and the momentum $\pi_{\phi}$. In fact, according to our comments, this contribution will be a quantum correction to a term that is not only quadratic in the perturbations, but in addition is proportional to the derivative of the matter field potential. Thus, for practical purposes, one would be allowed to neglect it.

Another result that is straightforward to obtain from our discussion is the difference, owing to choices of factor ordering, between the quantum constraint $\hat {\mathcal C}^{\mathrm d}$ which leads to an evolution equation of the dressed metric type and the quantum constraint $\hat {\mathcal C}$ of the preceding section. Using the same algebraic symmetrization for the products of $\hat {\mathcal H}_0^{-1}$ with the MS contributions in the two factors of the constraint $\hat {\mathcal C}^{\mathrm d}$, ignoring quantization prescriptions for the MS Hamiltonian and $\hat {\mathcal H}_0$, and recalling that $\hat \pi_{\phi}=-i\partial_{\phi}$, we get
\begin{equation}
\hat {\mathcal C}-\hat {\mathcal C}^{\mathrm d}= \Big[\hat \pi_{\phi},\hat {\mathcal H}_0+\frac{1}{2}\hat {\mathcal H}_0^{-1/2}\Big(\hat \Theta_{\mathrm e}+ \big(\hat \Theta_{\mathrm o} \hat \pi_{\phi}\big)_S \Big)\hat {\mathcal H}_0^{-1/2}\Big]-\frac{1}{2}\Big[
\hat {\mathcal H}_0^{-1/2},\Big[\hat {\mathcal H}_0^{1/2},\hat \Theta_{\mathrm e}+ \big(\hat \Theta_{\mathrm o} \hat \pi_{\phi}\big)_S \Big]\Big].
\end{equation}
This expression shows that the difference between the two constraints is equal to commutators between operators, and hence amounts to a choice of factor ordering. In this sense, we can say that the dressed metric approach may be related to a symplectic description of the perturbed FRW universes as a constrained system. Obviously, if one further truncates the formalism to remove all corrections to the zero modes quadratic in the perturbations, the symplectic canonical structure is lost, and the constraint no longer persists, since it modifies the dynamics of those modes precisely with quadratic perturbative contributions \cite{bojogianlufren}.

\section{Effective equations for the Mukhanov-Sasaki variables}\label{MSequations}

In this section, we will provide the effective equations for the MS variables in the quantization schemes that we have been discussing, extrapolating the experience gained in homogeneous models and assuming a direct relation between the annihilation and creation operators for the inhomogeneities and their classical counterpart. Let us start with the hybrid approach in the description of the perturbations obtained with the BO ansatz. In this case, the evolution of the perturbations is ruled by Eq. \eqref{constrainBO} [and Eq. \eqref{totalder}], which can be interpreted as the result of a constraint $\hat{\mathcal C}_{\mathrm{per}}$ that arises from the original constraint operator $\hat{\mathcal C}$ and is imposed on the sector of the model composed by the homogeneous degrees of freedom of the scalar field and the inhomogeneous modes, namely, on $\mathcal H_\mathrm{kin}^\mathrm{matt}\otimes\mathcal F$. Taking into account the densitization of the constraint [set in Eqs. \eqref{eq:C_0} and \eqref{quadhamMS}] and the definition of the homogeneous part of the lapse function, it is not difficult to realize that $\hat{\mathcal C}_{\mathrm{per}}/2$ generates evolution in a time $\bar T$ that, at leading perturbative order, is related with the proper one by $dt=V d\bar T$ (the factor of $1/2$ in the constraint is introduced here for later convenience). Assuming the validity of our condition a) of Sec. \ref{BOa}, this constraint on the wave function $\psi$ of the perturbations takes the form
\begin{equation}
\hat{\mathcal C}_{\mathrm{per}}=\hat \pi_{\phi}^2+ D_{\chi}(\phi) \hat \pi_{\phi} + E_{\chi}(\phi) - \Big\langle \hat{\Theta}_{\mathrm e}+ \big(\hat{\Theta}_{\mathrm o} \hat{\mathcal H}_{0}\big)_S- \frac{i}{2}{\mathrm d}_{\phi}\hat{\Theta}_{\mathrm o} \Big\rangle_{\chi} .
\end{equation}
Here, $D_{\chi}$ and $E_{\chi}$ are two functions of $\phi$ which depend on the state $\chi$ of the homogeneous geometry, and which we do not specify because they will not be important for our calculations.

According to our assumptions, the effective equations for the MS variables may then be computed using as evolution generator in the time $\bar T$ the effective constraint ${\mathcal C}_{\mathrm{per}}/2$ obtained by replacing $\hat \pi_{\phi}$ and the annihilation and creation operators for the inhomogeneities with their classical analogues, and taking standard Poisson brackets in the sector of homogeneous scalar field variables and inhomogeneous modes. Recalling expressions \eqref{thetaope}, we see that all the dependence of the evolution generator on $\pi_{{\tilde v}_{\vec n,\epsilon}}$ is given by a term $\langle \widehat{\left[1/V\right]}\,^{-2/3}\rangle_{\chi}\pi_{{\tilde v}_{\vec n,\epsilon}}^2/2$ coming from $\langle \hat{\Theta}_{\mathrm e}\rangle_{\chi}$. It is then most convenient to make a change of time from $\bar T$ to a time $\eta_{\chi}$ defined as
\begin{equation}\label{confotim}
d \eta_{\chi}= \langle \widehat{\left[1/V\right]}\,^{-2/3}\rangle_{\chi} d\bar T.
\end{equation}
Then, we straightforwardly get that $d_{\eta_{\chi}}{\tilde v}_{\vec n,\epsilon}=\pi_{{\tilde v}_{\vec n,\epsilon}}$, where $d_{\eta_{\chi}}$ denotes the derivative with respect to $\eta_{\chi}$.

Note that, with our definition, the time derivative $d\eta_{\chi}/d\bar T$ is strictly nonnegative (the operator $\widehat{\left[1/V\right]}$ is strictly positive in the orthogonal complement of the zero-volume state, where we have carried out our quantization), ensuring that the change of time is well defined. This time derivative is a function of only $\phi$ which, when evaluated on solutions to the effective equations, provides a time function. It is worth emphasizing that we could not have defined a change of time parameter had this time derivative been an operator. Hence, the expectation value on $\chi$ is essential in order to introduce the above change of time. We also point out that the change is state dependent, and hence the properties of the evolution in the times $\bar T$ and $\eta_{\chi}$ can be quite different when considered in the physical Hilbert space of the system. Finally, we notice the relation $d \eta_{\chi}= \langle \widehat{\left[1/V\right]}\,^{-2/3}\rangle_{\chi} dt/V$, and recalling that $V^{1/3}= l_0 \sigma e^{\tilde \alpha}$ is the scale factor (up to a multiplicative constant), we conclude that the new time can be interpreted in fact as a conformal time.

In order to get the effective MS equations, we still need to find the time derivative of the momentum variables $\pi_{{\tilde v}_{\vec n,\epsilon}}$, each of them obtained as the Poisson bracket of the variable with ${\mathcal C}_{\mathrm{per}}/2$ and divided by $\langle \widehat{\left[1/V\right]}\,^{-2/3}\rangle_{\chi}$. Defining
\begin{subequations}\label{varthetas}
\begin{align}
\big\langle \hat{\vartheta}_{\mathrm e,(\tilde v)}\big\rangle_{\chi}{\tilde v}_{\vec n,\epsilon}^2 &= -\frac{1}{\langle \widehat{[1/V]}\,^{-2/3}\rangle_{\chi}} \langle \hat{\Theta}_{\mathrm e}^{\vec{n},\epsilon}\rangle_{\chi}  - \tilde \omega_n^2 {\tilde v}_{\vec n,\epsilon}^2
- \pi_{{\tilde v}_{\vec n,\epsilon}}^2 ,\\
\big\langle \hat{\vartheta}_{\mathrm o,(\tilde v)}\big\rangle_{\chi}{\tilde v}_{\vec n,\epsilon}^2 &= -\frac{1}{\langle \widehat{[1/V]}\,^{-2/3}\rangle_{\chi}}\big\langle \big(\hat{\Theta}_{\mathrm o}^{\vec{n},\epsilon} \hat{\mathcal H}_{0}\big)_S- \frac{i}{2}{\mathrm d}_{\phi}\hat{\Theta}_{\mathrm o}^{\vec{n},\epsilon} \big\rangle_{\chi},
\end{align}
\end{subequations}
with the annihilation and creation-like variables in the above theta-operators treated as classical, we obtain
\begin{equation}\label{MSeqhybrid}
d^2_{\eta_{\chi}}{\tilde v}_{\vec n,\epsilon}=- {\tilde v}_{\vec n,\epsilon} \left[\tilde \omega_n^2 +
\big\langle \hat{\vartheta}_{\mathrm e,(\tilde v)}+ \hat{\vartheta}_{\mathrm o,(\tilde v)}\big\rangle_{\chi}\right].
\end{equation}

A number of comments are in order. First note that, from our definitions, the last factor in the square brackets of this MS equation is a function of only $\phi$, and hence of time when the scalar field is evaluated on the solutions to the effective equations. This factor contains quantum modifications with respect to the standard MS equation. Even so, the derived equations are still of harmonic oscillator type with time dependent frequencies. Besides, no dissipation term appears and the equations are hyperbolic in the ultraviolet regime, where $\tilde \omega_n^2$ dominates in the square brackets.

Using Eqs. \eqref{thetaope} and \eqref{varthetas}, and with our quantization prescriptions, we explicitly have that
\begin{subequations}\label{evaluate}
\begin{align}
\big\langle \hat{\vartheta}_{\mathrm e,(\tilde v)} \big\rangle_{\chi} &=
\frac{4\pi G}{3\big\langle \widehat{[1/V]}\,^{-2/3} \big \rangle_{\chi}} \big\langle\widehat{[1/V]}\,^{1/3}  \left(19 \hat {\mathcal H}_0^2 - 24 \pi G \gamma^2
\hat {\mathcal H}_0^2 \hat \Omega_0^{-2} \hat {\mathcal H}_0^2\right) \widehat{[1/V]}\,^{1/3}\big \rangle_{\chi}\nonumber\\
&+\frac{\big\langle \widehat{[1/V]}\,^{-2/3} \hat V^{2/3}\big \rangle_{\chi}}{\big\langle \widehat{[1/V]}\,^{-2/3} \big \rangle_{\chi}} \left(W^{\prime\prime}-\frac{16 \pi G}{3} W \right) ,\\
\big\langle \hat{\vartheta}_{\mathrm o,(\tilde v)} \big\rangle_{\chi} &=\frac{16 \pi G \gamma}{\big\langle \widehat{[1/V]}\,^{-2/3} \big \rangle_{\chi}}\big\langle \widehat{[1/V]}\,^{-1/3} \hat V^{1/3} |\hat \Omega_0|^{-1} \hat \Lambda_0 |\hat \Omega_0|^{-1} \hat V^{1/3} \widehat{[1/V]}\,^{-1/3} \big(\hat {\mathcal H}_0  W^{\prime} - \frac{i}{2} W^{\prime\prime}\big)\big \rangle_{\chi},
\end{align}
\end{subequations}
where $W$ is the matter field potential: $m^2\phi^2/2$ in our case. We have included the contribution of $\big\langle \hat{\vartheta}_{\mathrm o,(\tilde v)} \big\rangle_{\chi}$, although it contains only derivatives of the potential and, in view of our discussion in previous sections, we expect it to be negligible in practice.

It is reassuring that one would have arrived at the same result starting from the quantum constraint $\hat {\mathcal C}$ on the total kinematical Hilbert space of the system without introducing the BO approximation, by extrapolating the conjectures of LQC about the effective dynamics and with certain subtleties about the evaluation of the different terms of the homogeneous variables on effective solutions. Based on this extrapolation, one may accept that the evolution in the time $\bar T$ is generated by the effective constraint that one obtains by replacing in $\hat {\mathcal C}$ the annihilation and creation operators for the MS inhomogeneous modes again with their classical correspondents, the operators $\hat V$ and $\hat \pi_{\phi}$ (and $\hat \phi$) with their classical analogues as well, except for the mentioned caveats that we will comment on below, and the operators $\hat \Omega_0^2$ and $\hat \Lambda_0$ with the effective quantities $V^2 \sin^2{b}/\Delta$ and $\sgn{(v)}  V \sin{(2b)}/(2\sqrt{\Delta})$, respectively, where $b=\sqrt{\Delta} |V|^{-1/3} c$. Recall that $b$ is (up to a constant multiplicative factor) canonically conjugate to $V$ under Poisson brackets. As for the operator $\widehat{[1/V]}$, we also recall that it commutes with the volume operator, and hence can be expressed as a function of the latter using its spectral decomposition (see, e.g., \cite{galindo}). One can then find the equations of motion satisfied by the MS variables in a way similar to what we did in the BO scenario. The subtleties appear when one considers the different factors in these equations which depend on the homogeneous variables. In principle, those factors must be evaluated on an effective solution: precisely the solution on which the quantum state that admits the effective description is highly peaked. If the state is so peaked in a trajectory that, as far as the factors of the homogeneous variables are concerned, their evaluation in expectation values of the basic operators is essentially equal to the expectation values of those factors treated as operators, the way chosen to make the evaluation among these possibilities is irrelevant. If, on the other hand, there exist differences depending on how this evaluation is performed (something that would be the case if one considered generic functions on the homogeneous sector of the phase space), it is clear that, in order to recover the same results as in the BO ansatz, the prescription for the evaluation has to become that given in Eqs. \eqref{evaluate}. The same line of reasoning applies to the definition of the conformal time $\eta_{\chi}$ on the effective solution. With these remarks, the extrapolation of the effective dynamics found in LQC for homogeneous and isotropic systems seems to be valid in the present description of cosmological perturbations.

Finally, let us consider the effective equations that would follow from the description of Schr\"odinger type derived with the alternate factor ordering presented in Sec. \ref{orderin}. Recall that, apart from some issues related with the scaling of the inhomogeneities and the details of the quantization prescription, this description provides evolution equations for the perturbations similar to those obtained with the dressed metric approach. The generator of the evolution in the time $\phi$ is, according to Eq. \eqref{AANconstrain2}:
\begin{equation}\label{gen}
\frac{1}{2}\big\langle \hat {\mathcal H}_0^{-1/2}\left(\hat \Theta_{\mathrm e}+\big(\hat \Theta_{\mathrm o} \hat {\mathcal H}_0\big)_S\right) \hat {\mathcal H}_0^{-1/2} -\frac{i}{2}{\mathcal H}_0^{-1/2} {\mathrm d}_{\phi} \big(\hat \Theta_{\mathrm o}  \hat {\mathcal H}_0^{-1}\big) \hat {\mathcal H}_0^{1/2}\rangle_{\chi}.
\end{equation}

As in the above discussion of the effective equations in the hybrid approach, it is convenient to introduce a change of time, which will be determined by a function of $\phi$ (and hence of the original time) dependent on the state $\chi$ considered for the homogeneous geometry. We call this time $\eta_{\chi}^{\mathrm d}$, and define it through the relation
\begin{equation}
d\eta_{\chi}^{\mathrm d}  = \big\langle \hat {\mathcal H}_0^{-1/2} \widehat{\left[1/V\right]}\,^{-2/3} \hat {\mathcal H}_0^{-1/2}\big\rangle_{\chi} d\phi.
\end{equation}
We notice that the function of $\phi$ that determines the derivative $d\eta_{\chi}^{\mathrm d}/d\phi$ is strictly positive.

As an aside, note that the homogeneous scalar field $\phi$ and the time $\bar T$ that we introduced above are related on effective solutions by the evolution equation $d\phi/d\bar T=\pi_{\phi}$. In the effective description, we may use this relation to change times, replacing the momentum $\pi_{\phi}$ by its value on the considered solution, which at dominant order in the perturbations coincides with the expectation value of $\hat {\mathcal H}_0$ on $\chi$. In this sense, one would obtain
\begin{equation}
d\eta_{\chi}^{\mathrm d} = \big\langle \hat {\mathcal H}_0^{-1/2} \widehat{\left[1/V\right]}\,^{-2/3} \hat {\mathcal H}_0^{-1/2}\big\rangle_{\chi} \big\langle \hat {\mathcal H}_0\big\rangle_{\chi} \, d\bar T .
\end{equation}
Comparing this relation with Eq. \eqref{confotim}, we see that $\eta_{\chi}^{\mathrm d}$ can be interpreted again as a conformal time, and that its definition corresponds to a different recipe for the evaluation of the homogeneous scale factor.

Employing the generator \eqref{gen} for the evolution in the time $\phi$ (under Poisson brackets and, again, with the annihilation and creation-like variables regarded as classical), the introduced change of time to the conformal one $\eta_{\chi}^{\mathrm d}$, and a calculation similar to that explained above for the BO ansatz in the hybrid approach, one easily concludes that the effective MS equation adopts now the form
\begin{equation}\label{MSeqdres}
d^2_{\eta^{\mathrm d}_{\chi}}{\tilde v}_{\vec n,\epsilon}=- {\tilde v}_{\vec n,\epsilon} \left[\tilde \omega_n^2 +
\big\langle \hat{\vartheta}^{\mathrm d}_{\mathrm e,(\tilde v)}+ \hat{\vartheta}^{\mathrm d}_{\mathrm o,(\tilde v)}\big\rangle_{\chi}\right],
\end{equation}
where
\begin{subequations}\label{varthetasdres}
\begin{align}
\big\langle \hat{\vartheta}^{\mathrm d}_{\mathrm e,(\tilde v)}\big\rangle_{\chi}{\tilde v}_{\vec n,\epsilon}^2 &= -\frac{1}{\langle \hat {\mathcal H}_0^{-1/2}\widehat{[1/V]}\,^{-2/3}\hat {\mathcal H}_0^{-1/2}\rangle_{\chi}} \langle \hat {\mathcal H}_0^{-1/2}\hat{\Theta}_{\mathrm e}^{\vec{n},\epsilon}\hat {\mathcal H}_0^{-1/2}\rangle_{\chi}  - \pi_{{\tilde v}_{\vec n,\epsilon}}^2 - \tilde \omega_n^2 {\tilde v}_{\vec n,\epsilon}^2,\\
\big\langle \hat{\vartheta}^{\mathrm d}_{\mathrm o,(\tilde v)}\big\rangle_{\chi}{\tilde v}_{\vec n,\epsilon}^2 &=-\frac{1}{\langle \hat {\mathcal H}_0^{-1/2}\widehat{[1/V]}\,^{-2/3}\hat {\mathcal H}_0^{-1/2}\rangle_{\chi}} \nonumber\\
&\times\Big\langle \hat {\mathcal H}_0^{-1/2}\big(\hat{\Theta}_{\mathrm o}^{\vec{n},\epsilon} \hat{\mathcal H}_{0}\big)_S \hat {\mathcal H}_0^{-1/2} -\frac{i}{2} \hat {\mathcal H}_0^{-1/2} {\mathrm d}_{\phi} \big(\hat \Theta_{\mathrm o}^{\vec{n},\epsilon} \hat {\mathcal H}_0^{-1}\big) \hat {\mathcal H}_0^{1/2}\Big\rangle_{\chi},
\end{align}
\end{subequations}
with the convention that the variables of the inhomogeneous modes are treated classically. The parallelism with Eqs. \eqref{varthetas} and \eqref{MSeqhybrid} is evident. Except for a contribution that (appropriately rewritten) is proportional to $\langle\hat {\mathcal H}_0^{-1/2}  \hat \Theta_{\mathrm o}^{\vec{n},\epsilon} {\mathrm d}_{\phi} \hat {\mathcal H}_0^{-1/2} \rangle_{\chi}$ and which may be attributed to a specific choice of factor ordering ---moreover, which is negligible if the derivative of the potential is ignorable---, the difference between the two effective MS equations can be described by saying that, in the ratios of expectation values, the state $\chi$ is replaced in the present case with the state $\hat {\mathcal H}_0^{-1/2} \chi$. If the state is so highly peaked that the expectation value of products of operators coincides with the product of the corresponding expectation values, then no discrepancy is expected if the same prescriptions are adopted to quantize the quadratic contributions of the inhomogeneities (together with ${\mathcal H}_0$ and the inverse-volume operator) as before.

\section{Conclusions}\label{conclusions}

We have discussed the hybrid quantization approach to the treatment of cosmological perturbations around flat homogeneous and isotropic universes containing a minimally coupled scalar matter field in the framework of LQC and employing MS gauge invariants. The use of MS variables clarifies the independence of the results with respect to (perturbative) gauge transformations. Moreover, it can be considered as a first step towards a formulation of the perturbations entirely in terms of gauge invariants, linear perturbative constraints, and appropriate momenta. Such a  description, when completed into a canonical transformation in the whole phase space of the system, would allow one to reach a quantization with no gauge fixing. In this quantization, one might analyze directly the closure of the entire algebra of constraints, hence providing links with the so-called {\it effective approach} to the description of cosmological perturbations. In addition, the use of MS variables facilitates the comparison of the procedures and results of the hybrid approach with those corresponding to the dressed metric approach.

The hybrid approach is based on two approximations. On the one hand, the validity of the hybrid hierarchy, in which the effects of the loop quantum geometry on the inhomogeneous modes are neglected against their influence on the homogeneous degrees of freedom. On the other hand, the validity of a perturbative truncation of the action at quadratic order, with perturbations describing inhomogeneities. This truncation permits that the system remains as a constrained, symplectic one, as it is typical in gravitational systems. In turn, this permits the quantum treatment of the spacetime structures, including the metric, since it makes possible a genuine quantization of the perturbed metric, rather than describing the perturbations as test fields over a metric that is quantum corrected. In this latter situation, found in the dressed metric approach, one is bound to a QFT on a quantum/effective curved background, instead of facing a genuine quantum theory of a cosmological system (even if this system is constructed with some approximations).

With the above hybrid and truncation schemes, we have reformulated the cosmological model described previously in Ref. \cite{hybridflat} in terms of MS variables, determining canonical momenta for them in the inhomogeneous sector of the system and completing this change of inhomogeneous variables into a canonical transformation in the whole of the phase space, including homogeneous degrees of freedom. We have also calculated the corresponding MS Hamiltonian, providing the quadratic contribution in inhomogeneities to the zero mode of the Hamiltonian constraint of the entire system. This constraint can be easily found using the introduced canonical transformation, starting from the total Hamiltonian constraint and respecting our quadratic truncation. Alternatively, it can be computed by regarding our change of inhomogeneous variables as a background dependent one, and finding the new Hamiltonian for the inhomogeneous perturbations with the standard formulas for canonical transformations that are explicitly time dependent. Both methods lead to the same result.

We then revisited the hybrid quantization of the system, where only the zero mode of the Hamiltonian constraint remains to be imposed \`a la Dirac. Furthermore, we focused our attention on states in which the dependence on the homogeneous degrees of freedom can be separated from that on the inhomogeneous perturbations, treating the homogeneous part of the scalar field as an internal time, inspired by the BO ansatz of atomic physics. With this ansatz, and neglecting transitions in the quantum state of the homogeneous geometry sector mediated by the constraint, we have arrived at a kind of constraint equation on (the part of) the wave function of the perturbations. This equation is a second-order one in the intrinsic time. Under certain hypotheses, which can be checked in each specific case under consideration, this evolution equation can be approximated by one of  Schr\"odinger type, which resembles the evolution equation obtained in the dressed metric approach. In particular, in practical situations, the expectation value of the momentum of the homogeneous scalar field should not be numerically of the same order as the quadratic perturbations; otherwise the perturbation scheme, although consistent in an abstract asymptotic limit, should not be expected to lead to a good approximation.

In addition to all this, we have also proceeded to quantize the system with an alternate factor ordering, still within the lines of the hybrid approach, and hence maintaining the description of the model as a constrained symplectic manifold. This alternate procedure has been motivated by the fact that the terms that one needs to neglect in the usual choice of quantum representation of the constraint in the hybrid approach, in order to obtain a Schr\"odinger equation for the perturbations similar to that of the dressed metric formalism, can be realized as ambiguities in factor ordering. Then, we have proven that there exists a factor ordering that leads to similar results as a deparametrization of the system in terms of the internal time $\phi$. Introducing again a BO ansatz and neglecting as well quantum transitions in the sector of the FRW geometry, we have obtained an evolution equation for the inhomogeneities that is the parallel of the equation deduced in Refs. \cite{AAN1,AAN2}, except in what concerns a different scaling of the MS variables (necessary if one wants a unitary dynamics in the regime of QFT in curved backgrounds) and some issues about the quantization prescriptions for the homogeneous degrees of freedom. In this specific sense, one can say that a formulation like that of the dressed metric approach can be derived from the hybrid approach with a particular choice of factor ordering and of prescriptions in the construction of the quantum representation. As we have pointed out, starting from the hybrid approach, one has at one's disposal a symplectic manifold description, a constrained dynamics arising from the constraints of general relativity, and a concept of quantum metric. If one insists on keeping only linear perturbations to all the metric degrees of freedom (a truncation which differs from that at quadratic order in the action), one has to renounce to the canonical symplectic structure, the constraint is not longer satisfied in the total system, and the perturbations evolve indeed as test fields, missing a genuine quantum spacetime structure.

We have also discussed the effective equations for the MS variables associated with our hybrid quantization, included the case with the proposed alternate factor ordering, assuming a direct replacement of the annihilation and creation operators of the inhomogeneities with their classical counterparts. We have seen that the BO ansatz sheds light on the evaluation of the homogeneous geometry factors of the effective MS equation in the hybrid approach, identifying these factors with expectation values of operators in the corresponding quantum state. When this state is sufficiently peaked, the expectation values may well reproduce the values on the peak trajectory, but the derivation is valid in more general cases. In addition, we have seen that the effective MS equations of the hybrid approach do not suffer fundamental changes when one switches to the alternate factor ordering related with the dressed metric approach. The effective MS equation is of second order in an adequately defined conformal time, whose definition changes slightly with the adopted factor ordering and depends on the particular quantum state considered for the homogeneous geometry. This second-order equation is hyperbolic in the ultraviolet sector, no dissipative term appears, and it is only the effective time dependent potential that is altered with the alternate factor ordering. This effective MS equation supplies the information needed to compute the modified power spectrum of primordial perturbations in the cosmic background radiation. A similar analysis can be carried out in the case of tensor perturbations, whose treatment is even easier owing to the simpler potential in the corresponding MS Hamiltonian.  Finally, it is worth commenting that, although the effective MS equations that we have derived remain hyperbolic for modes of asymptotically large frequency, the actual Lorentzian or Euclidean character of the geometry in an effective description should be studied carefully from the consideration of the quantum metric, where the homogeneous degrees of freedom have been corrected with quadratic terms in the inhomogeneous perturbations in order to keep our truncation of the action at quadratic perturbative order. Although, in principle, the effects of these corrections ---and of possible changes of lapse associated with redefinitions of the constraints at the considered perturbative order--- should not affect the global character of the spacetime metric in a way independent of the perturbations, further investigation seems necessary to have a better understanding of this issue. This study will be the subject of future research.

\acknowledgments{We would like to thank G. Calcagni, J. Cortez, M. Mart\'{\i}n-Benito, D. Mart\'\i{}n-de Blas, S. Tsujikawa, and J.M. Velhinho for discussions. G. Mena acknowledges I. Agull\'o and M. Bojowald for correspondence about their viewpoints on the quantization of cosmological perturbations in LQC; some of their ideas coincide with those reflected in this article since the first draft versions. This work was supported by the Projects No.\ MICINN/MINECO FIS2011-30145-C03-02 from Spain and i-Link0484 of the i-Link cooperation program of CSIC. M.F.-M. acknowledges CSIC and the European Social Fund for financial support under the grant JAEPre\_2010\_01544. J.O. acknowledges Pedeciba.}

\bibliographystyle{plain}

\end{document}